\documentclass[fleqn,usenatbib]{mnras}

\usepackage{newtxtext,newtxmath}
\usepackage{xcolor}

\usepackage[T1]{fontenc}
\usepackage{array}
\usepackage{caption} 

\usepackage{needspace}

\DeclareRobustCommand{\VAN}[3]{#2}
\let\VANthebibliography\thebibliography
\def\thebibliography{\DeclareRobustCommand{\VAN}[3]{##3}\VANthebibliography}

\usepackage{graphicx}	
\usepackage{amsmath}	

\title[Caught in the heat]{GATOS XI : Excess dust heating in the Narrow Line Regions of nearby AGN revealed with JWST/MIRI }

\author[Haidar et al.]{
Houda Haidar,$^{1}$\thanks{E-mail: h.haidar2@newcastle.ac.uk}
David J. Rosario,$^{1}$
Ismael Garc\'{\i}a\mbox{-}Bernete,$^{2}$
Almudena Alonso\mbox{-}Herrero,$^{2}$
Anelise Audibert,$^{3,4}$
\newauthor
Steph Campbell,$^{1}$
Chris M. Harrison,$^{1}$
Tiago Costa,$^{1}$
Laura Hermosa Mu\~noz,$^{2}$
Fran\c{c}oise Combes,$^{5}$
\newauthor
Dimitra Rigopoulou,$^{6}$
Claudio Ricci,$^{7,8}$
Cristina Ramos Almeida,$^{3,4}$
Enrica Bellocchi,$^{9,10}$
Peter Boorman,$^{11,12}$
\newauthor
Andrew Bunker,$^{6}$
Richard Davies,$^{13}$
Daniel Delaney,$^{14,15}$
Tanio D\'{\i}az Santos,$^{16,17}$
Federico Esposito,$^{18}$
\newauthor
Victoria A. Fawcett,$^{1}$
Poshak Gandhi,$^{19}$
Santiago Garc\'{\i}a\mbox{-}Burillo,$^{18}$
Omaira Gonz\'alez\mbox{-}Mart\'{\i}n,$^{20}$
\newauthor
Erin K. S. Hicks,$^{14,15,21}$
Sebastian F. H\"onig,$^{19}$
Alvaro Labiano,$^{22}$
Nancy A. Levenson,$^{23}$
\newauthor
Enrique Lopez\mbox{-}Rodriguez,$^{24,25}$
Chris Packham,$^{21,26}$
Miguel Pereira\mbox{-}Santaella,$^{27}$
Rogemar A. Riffel,$^{28,29}$
\newauthor
Alberto Rodr\'{\i}guez Ardila,$^{30,31}$
John Schneider,$^{21}$
T. Taro Shimizu,$^{13}$
Marko Stalevski,$^{32,33}$
\newauthor
Montserrat Villar Mart\'{\i}n,$^{28}$
Martin Ward,$^{34}$ 
Lulu Zhang,$^{21}$ Gillian Leeds,$^{24}$ Fergus R. Donnan$^{35}$  \\
\hfill \\
\noindent (Affiliations are listed at the end of the paper)}

\date{Accepted XXX. Received YYY; in original form ZZZ}

\pubyear{\the\year{}}

\begin{document}
\label{firstpage}
\pagerange{\pageref{firstpage}--\pageref{lastpage}}
\maketitle

\begin{abstract} 
We present JWST/MIRI imaging  of eight nearby Active Galactic Nuclei (AGN) from the GATOS survey to investigate the physical conditions  of extended dust in their narrow line regions (NLRs). In four galaxies (ESO 428$-$G14, NGC 4388, NGC 3081, and NGC 5728), we detect spatially resolved dust structures 
extending  $\sim$100-200 pc along the NLR.
In these systems,  we find a strong link between the morphology of the dust, the radio ejecta, and the coronal [\ion{Si}{VI}] emission, implying that dust carries imprints of the processes shaping the NLR.  
Using spatially resolved spectral energy distributions, we show that dust in the NLR has systematically steeper slopes than star forming clumps.  This dust emits at temperatures in the range $150- 220\, \rm K$, at a distance of $\sim$150 pc from the nucleus.
Using simple models, we show that, even under optimistic assumptions of grain size and AGN luminosity, the excess MIR emission cannot be explained by AGN illumination alone. We interpret this excess heating as in-situ. We show that shocks with velocities of $v_{\rm shock} \sim 200- 400 \, \rm km/s$ in dense gas can close this gap, and in some cases even account for the total observed emission.
This, combined with multiple lines of evidence for shocks in these regions, supports a scenario in which shocks not only coexist with dust but may be playing a key role in heating it. Our findings reveal shocks may be an important and previously overlooked driver of extended dust emission in the central hundreds of parsecs in AGN.
\end{abstract}




\section{introduction}



Interstellar dust lies at the heart of
galaxy evolution,
serving as a key driver of the thermodynamics and chemistry of galaxies \citep[e.g.][]{Salpeter+77, Whittet+93, WeingartnerDraine2001, Draine2003, Draine+11}.
Dust shapes the spectral energy distributions (SEDs) of galaxies, 
by absorbing and scattering optical and ultraviolet (UV) photons, then re-emitting this energy predominantly as infrared (IR) radiation 
\citep[e.g.][]{Draine1984,1985ApJ...292..494D,Li-Draine-2001,2001PASP..113.1449C,Tielens+2005}.
Additionally, grain surfaces serve as a medium where essential chemical reactions form molecules \citep[e.g.][]{Hollenbach1971}.
At the same time, these grains also act as the primary reservoir in which refractory elements 
are stored, while regulating their availability in the gas phase through accretion and destruction processes, giving rise to a diverse and complex chemistry \citep[e.g.][]{Whittet+93,WeingartnerDraine2001}. \\

Dust proves especially important in the context of Active Galactic Nuclei (AGN). These luminous systems, powered by the accretion onto supermassive black holes (SMBHs), are responsible for regulating star formation in their host galaxies through feedback processes driven by  powerful winds and jets \citep[e.g.][]{1993MNRAS.263..323T,1998A&A...331L...1S,2012ARA&A..50..455F,2015ARA&A..53..115K,2024Galax..12...17H}. 
In many AGN, strong optical extinction has been attributed to the presence of an optically obscuring thick structure, dubbed as ``torus'' \citep[e.g.][]{1988ApJ...329..702K,2004Natur.429...47J,2008ApJ...685..160N,RamosAlmeida2017}. In the AGN unification model, this dusty torus has been held responsible for the dichotomy between type 1 and type 2 AGN \citep{Antonucci+85,Urry+95}. The dusty torus has now been  imaged in the submillimeter with ALMA\footnote{Atacama Large Millimeter/submillimeter Array (ALMA)} and in the near- and mid- infrared with VLTI\footnote{Very Large Telescope Interferometer} with spatial resolutions spanning from few to tens of parsecs \citep[][]{GarciaBurillo+16,GarciaBurillo+19,GarciaBurillo+21,Gallimore+16,Alonso-Herrero+18,Alonso-Herrero+19,Alonso-Herrero+23,Combes+19,GRAVITY+20,Gamez+22}. \\

In addition to this, substantial effort in high-resolution infrared astronomy over the past decades has revealed compelling evidence of a significant amount of dust located beyond the torus, and extending from a few parsecs to tens of parsecs along the narrow line region (NLR) in a number of nearby AGN \citep[e.g.][]{Honig+12,Honig+13,Tristram+14,Leftly2018}.
In some cases, this has been interpreted as a dusty wind originating from the inner edges of the torus \citep[see][for a review]{2019ApJ...884..171H}. For dust to be launched from the torus itself,
radiation pressure must exceed gravitational forces, as theoretically demonstrated by \cite{2006MNRAS.373L..16F} and also \citet{2020ApJ...900..174V}. \\

The presence of dust along outflowing regions in Seyferts has long been recognised \citep[e.g.][]{2000AJ....120.2904B,2003ApJ...587..117R}, with early studies already highlighting its role as a key component of the NLR \citep[e.g.][]{1985aagq.conf..259M,1989agna.book.....O}. The latter is subject to outflows, driven by winds or jets, which can alter the distribution of the embedded dusty gas. Dust can carry imprints of such processes, making it a direct tracer of how AGN impact their surroundings. While AGN photoionisation is considered the primary radiation source powering emission in the NLR, several works have shown that it alone may not be enough, and 
that contributions from shocks cannot be neglected  
\citep[e.g.][]{1989ApJ...339..689V, morse1996viability,1997ApJS..110..287F}. 
In many Seyferts, shocks have been shown to be a viable mechanism for shaping the spatial and velocity distribution of gas in the NLR, particularly in regions where the jet or wind interacts with the interstellar medium (ISM)
\citep[e.g.][]{1998ApJ...502..199F,1999ApJ...521..531E,2004AJ....127..606W,2005A&A...444L...9M,2006AJ....132..546G,2010MNRAS.408..565R,2010ApJ...711L..94R,2012ApJ...758...95M,2017MNRAS.472.3842W}.  \\


Recent JWST/MIRI imaging of the nearby Seyfert ESO 428$-$G14 has successfully resolved parsec-scale dust along the NLR \citep[][Fig. 1 therein]{Haidar+24}, extending up 100 pc from the nucleus. This dust bears a striking morphological resemblance to the radio and [\ion{Si}{VI}] emission, coinciding with regions where shocks are prevalent \citep[][]{1996ApJ...470L..31F, 2006MNRAS.373....2R, 2018MNRAS.481L.105M}. By comparing AGN illumination, dusty winds, and shocks as potential sources of dust heating, we argued in \citet{Haidar+24} that the radiative output from fast shocks can reproduce the observed dust temperatures better than heating from AGN radiation fields alone. This puts forward a scenario in which shocks play a central role in driving the infrared emission within the central few hundred parsecs of AGN. \\

\begin{table*}
    \caption{Sample Properties. }
    \label{tab:agn_properties}
    \centering
    \begin{tabular}{lccccccccc}
        \hline
        AGN & Type & Distance & Scale & $\log_{10}(L_{2\text{-}10\mathrm{\,keV}})$ & $\log_{10}(L_{\mathrm{bol}})$ & PA$_{\mathrm{NLR}}$ & PA$_{\mathrm{radio}}$ & PA$_{\mathrm{disc}}$ \\
         &  & (Mpc) & (pc/") & ($\mathrm{erg\,s}^{-1}$) & ($\mathrm{erg\,s}^{-1}$) &(deg) & (deg) & (deg) \\
        \hline\hline
        NGC~4388     &  Sy\,2      & 19.2 &   93    & 42.5 & 43.7 & 30   & 30   & 90 \\
        ESO~428--G14 &   Sy\,2     & 23.2 &  112 & 41.55 & 42.74 & 130  & 127  & 62 \\  
        NGC~3081     &   Sy\,2     & 37 &  180 & 42.78 & 43.98 & 164  & 164  & 60 \\
        NGC~5728     &   Sy\,2     & 39 &  190  & 42.83 & 44    & 120  & 127  & 53 \\
        \hline
        NGC~3227     &   Sy\,1.5   & 15 &   73    & 42.0 & 43.19 & 195  & 170  & 68 \\
        NGC~2992     &   Sy\,1.5$\sim$2 & 30.9 &  150 & 41.94 & 43.12 & 125  & 154  & 90 \\
        NGC~7172     &   Sy\,2     & 33.9 &  180  & 42.67 & 43.86 & 60   & 90   & 53 \\
        NGC~5135     &   Sy\,2     & 64.8 &  281   & 43.25 & 44.45 & 30   & --   & 25 \\
        \hline
    \end{tabular}
    \\
\small  \textit{Notes --} (1) Target name. (2) Type based on optical AGN classification from \citet{2014MNRAS.439.1648A}. (3) Redshift-independent distances are taken from NED. (4) The pixel scale is derived from the distance. (5) Hard X-ray luminosities (2–10 keV) are computed from the integrated Swift/BAT fluxes, unless stated otherwise (see Section~\ref{sec:bol}). (6) Bolometric luminosities are obtained by applying a bolometric correction to the X-ray luminosities (see Eq.~\ref{eq:bolcorr}). (7 \& 8) The NLR and radio position angles (PA) are taken from \citet{2016ApJ...822..109A}. (9) Disk inclinations are taken from HyperLeda.
\end{table*}

Finding dust to be co-spatial with shocked regions raises a critical question about its survival. It is well established that processes such as sputtering and shattering in shocks are the key drivers of dust destruction in the ISM \cite[e.g.][]{1994ApJ...433..797J,1994ApJ...431..321T,1996ApJ...469..740J}. Sputtering, in particular, releases refractory elements, such as iron (Fe), 
that were previously locked within dust grains, back into the gas phase. 
Since [\ion{Fe}{II}] is also an efficient coolant in shocked gas \citep[e.g.][]{2004ApJ...614L..69H}, enhanced [\ion{Fe}{II}] emission is commonly used as a signature of shocks \citep{1979ApJ...231..438D}.
Several hypotheses and models have been put forward to explain how
dust endures the passage of a shock: shielding by cool clouds, longer destruction timescale,
resistance through coagulation or consistent replenishment processes at play \citep[][]{2001MNRAS.328..848V,2024arXiv240303711R,2024arXiv240808026C,2025NewA..11402293D,2024MNRAS.528.5364K,2022MNRAS.509.3163P,2024arXiv240303711R}. Additionally, the detection of molecular outflows 
suggests
that dust may survive for similar reasons, such as shielding by dense clumps and rapid
reformation mechanisms \citep[][]{2017ApJ...850...40R, 2018MNRAS.474.3673R, 2018MNRAS.478.3100R, 2022MNRAS.510..551F, 2024MNRAS.528.5008O}. \\

Building on our pilot study of ESO 428$-$G14,  we expand this analysis to the full sample from the JWST Cycle 1 programme \textit{``Dust in the Wind''} (ID: 2064, PI: D. Rosario) within the ambit of the \textit{Galactic Activity, Torus, and Outflow Survey} (\href{https://gatos.myportfolio.com}{GATOS}) collaboration \citep[see also][]{GarciaBurillo+21,Alonso-Herrero2021,GarciaBernete+23,2024ApJ...974..195Z,2024A&A...691A.162G,2024arXiv241118738F}, which consists of eight nearby AGN (see Table~\ref{tab:agn_properties}) that show some evidence of extended dust from ground-based observations \citep[][]{2016ApJ...822..109A}. Our goal is to explore (1) the extent and morphology of resolved dust in the NLR (2) whether this dust co-exists with shocks, and (3) whether shock heating can contribute to the observed dust emission. 
This paper is organised as follows. The observations and data reduction are presented in Section~\ref{methods}. The results are described in Section~\ref{results} and discussed in Section~\ref{sec:discussion}. A summary of our results is given in Section~\ref{conclusion}.

\section{Observations \& METHODS}
\label{methods}

\subsection{\textit{JWST}/MIRI imaging}
\label{method:jwstimaging}

All targets (see Table~\ref{tab:agn_properties}) were observed with the JWST Mid-Infrared Instrument (MIRI) in five broadband imaging filters: F560W, F1000W, F1500W, F1800W, and  F2100W (corresponding to central wavelengths $\rm \lambda_{central}= 5.6,10,15,18,$ and 21 $\mu$m, respectively).  The angular resolution of the MIRI imaging is set by the point-spread function (PSF), with full widths at half maximum (FWHM) of $0.2''$, $0.32''$, $0.48''$, $0.59''$, and $0.67''$ for the filters listed above\footnote{See also: https://jwst-docs.stsci.edu/mid-infrared-instrument/miri-observing-modes/miri-imaging\#gsc.tab=0.}. For key morphological comparisons discussed later in Section~\ref{results}, we focus on the F1000W band, which corresponds to a physical PSF size of $\sim 23$--$90$ pc, depending on the target (see Table ~\ref{tab:agn_properties}). The SUB256 subarray was used ($28\farcs2$ on a side), along with a fast readout mode and the shortest allowed number of groups ($N_{\rm groups} = 5$) to minimise the saturation of the bright nuclear emission expected in the targets. A relatively large number of integrations ($N_{\rm ints} = 50$) and four dithered exposures yielded a final exposure time of 358 seconds in each filter. The raw data were downloaded from the Mikulski Archive for Space Telescopes (MAST)\footnote{https://mast.stsci.edu/portal/Mashup/Clients/Mast/Portal.html }, and processed using the JWST pipeline python package version 10.2 with CRDS reference files \texttt{jwst\_1097.pmap}. \\

Data reduction and subtraction of the  PSF was performed as in \cite{Haidar+24}. Given the bright nature of these galaxies in the MIR, most images had a few partially-saturated pixels at the nucleus. These specific pixels were restored from up-the-ramp readouts by turning off the first-frame flagging step during Stage 1 of the JWST pipeline \citep[see][for more details]{Haidar+24}. 
This enables the flux from the first group to be included, yielding two unsaturated groups from which a reliable ramp slope can be estimated\footnote{see also JWST Technical Report JWST-STScI-008866, SM-12}.

In the specific case of NGC 3227, only a single good group was available.  
To address this, we turned off the \texttt{suppress\_one\_group} parameter in the ramp-fitting step of Stage 1, thus permitting ramp estimates on a single group. Further details on target selection, data processing, and the removal of the bright central point-source emission will be provided in \textcolor{blue}{Rosario et al., in prep.} 
\\

\begin{figure*}
  \centering
  {\includegraphics[width=0.7\textwidth]{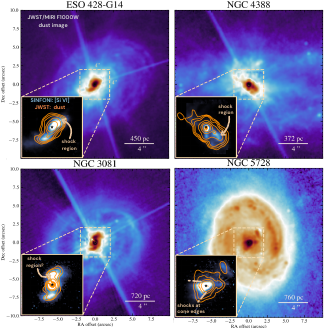}}
  \caption{
   JWST/MIRI F1000W images of the central regions of four galaxies (ESO 428$-$G14, NGC 4388, NGC 3081, and NGC 5728). North is up and East is to the left. The images have been PSF subtracted and decontaminated from emission lines. Each main panel shows the pure dust continuum emission in logarithmic scale, highlighting extended structures. The zoom-ins display the flux maps of [\ion{Si}{VI}] emission observed with SINFONI, matched to a $\sim 4'' \times 4''$ field of view. Overlaid contours show the extended dust emission as the main panels. Where relevant, arrows mark regions where shocks are prevalent, based on [\ion{Fe}{II}] enhancement.
}
  \label{dust_in_NLR_AGN}
\end{figure*}

\begin{figure*}
  \centering
  {\includegraphics[width=0.7\textwidth]{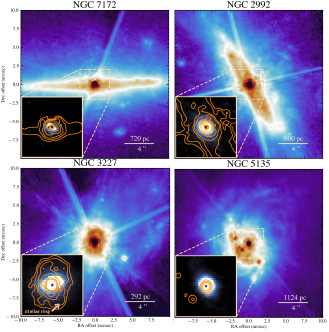}}
  \caption{ Similar to Fig.~\ref{dust_in_NLR_AGN} but for NGC 7172, NGC 2992, NGC 3227, and NGC 5135.}
  \label{NO_dust_in_NLR_AGN}
\end{figure*}

\subsection{VLT/SINFONI spectroscopy}
K-band ($\rm 1.9-2.5\, \mu m$) integral-field spectroscopic data is available for all our targets, obtained using the Spectrograph for INtegral Field Observations in the Near Infrared (SINFONI) on the European Southern Observatory (ESO) Very Large Telescope (VLT) \citep{Eisenhauer2003,Bonnet2004}. Data reduction was carried out with the SINFONI Data Reduction Software.

For most targets, the observations were obtained with the 100\,mas plate scale, except in the case of  NGC~3227 and NGC~5135, which were observed with the 250\,mas plate scale. After processing, the effective spatial sampling of the datacubes is half the platescale, corresponding to $0.05''\,$/pix for the K100 data (FoV $\sim 4''\!\times 4''$) and $0.125''\,$/pix for the K250 data (FoV $\sim 8''\!\times 8''$). All targets are obtained with adaptive optics (AO) assistance except in the case of NGC 3227 and NGC 5135. The AO-assisted targets achieve an average angular resolution of  PSF FWHM $\sim$ $0.17''$, while the non-AO observations have a typical PSF of $0.41''$. As we discuss later in Section ~\ref{results}, for the key targets exhibiting dusty NLRs (see Figs.~\ref{NO_dust_in_NLR_AGN} and ~\ref{fig:all_radio_dust} ), these SINFONI observations provide resolutions that are comparable to, if not superior to, JWST's F1000W filter (see Section~\ref{method:jwstimaging}).  \\

The emission line flux maps for [\ion{Si}{VI}]$\lambda 1.96~\mu$m  were generated using the custom IDL code LINEFIT \citep{davies2011}, which fits the emission line of interest with an unresolved line profile (a sky line) convolved with a Gaussian, as well as a linear function to the line-free continuum. This  single-component Gaussian  fitting procedure was performed for each spaxel of the data cube. More detail on the SINFONI reduction and analysis can be found in \citet{2025ApJ...984..163D}.

\subsection{VLA radio continuum imaging}
\label{sec:radio-method}
All radio maps for this study (except NGC 4388, for
which the radio map was kindly provided by \citet{2024ApJ...961..230S}) were extracted from the National Radio Astronomy Observatory (NRAO) Very Large Array (VLA) \href{https://science.nrao.edu/facilities/vla/archive/index}{Archive
 Survey (NVAS) Images portal}, which provides pipeline-processed radio continuum images for a major subset of the VLA archive up to 2005. We rely on the AIPS-processed pipeline outputs from NVAS, with no additional processing or cleaning. We select images of all targets with angular resolutions $<1''$ and sufficient sensitivity to detect extended radio emission. For each galaxy, we adopt the radio band showing the most prominent extended structure along the NLR. These are: 4–8.5 GHz for ESO 428–G14, NGC 4388, NGC 5135, NGC 3227, and NGC 2992; and 1.5 GHz for NGC 3081, NGC 5728, and NGC 7172.  The 4–8.5\,GHz observations provide higher angular resolution while the 1.5\,GHz data have lower resolutions but are  more sensitive to low surface brightness emission. This leads to the different spatial scales probed across the sample, as seen in Figs.~\ref{fig:all_radio_dust} and ~\ref{fig:dust_radio_no_coupling}. The angular resolution of the radio images, as measured from the beams, is $0.44''\times0.20''$ for ESO~428–G14, $0.38''\times 0.27''$ for NGC~4388, $2.25''\times 1.14''$ for NGC~3081, and $2.02''\times 1.16''$ for NGC~5728 (see middle panel in Fig.~\ref{fig:all_radio_dust},  bottom left for the beams). For ESO 428-G14 and NGC 4388, these radio maps provide sufficient resolution to compare with JWST's F1000W filter (PSF FWHM $\sim 0.32''$). For NGC~5728, while we cannot compare dust-radio morphologies, one could still constrain the orientation of the jet, as also presented in earlier studies \citep[see e.g.,][]{2019MNRAS.490.5860S}. In the case of NGC~3081, the radio map is unresolved at the available resolution and no morphological conclusions are drawn in this case, and the map is presented for completness only. We present more detail on this later in Section~\ref{sec:dustradiolink}. 
 
 For the additional galaxies shown in Appendix~\ref{ap:dust-radio-extra} (Fig.~\ref{fig:dust_radio_no_coupling}), the corresponding beam sizes are $1.3''\times 0.8''$ (NGC~5135), $3.5''\times 1.1''$ (NGC~7172),  $0.28''\times 0.24''$ (NGC~3227), and $0.52''\times 0.39''$ (NGC~2992).

\subsection{Bolometric luminosities}
\label{sec:bol}
To compute AGN bolometric luminosities ($L_{\rm bol, AGN}$), we use estimates of the absorption-corrected 2-10 keV X-ray luminosity from \citet{2017ApJS..233...17R}, in which X-ray spectral fits have been performed for all AGN in the 70-month \textit{Swift}/BAT all-sky survey. ESO 428$-$G14 and NGC 5135 are not covered by the survey, in which case, the X-ray luminosity is extracted from \citet{2006ApJ...648..111L} and \citet{2012MNRAS.419.2089S} respectively. 
The X-ray luminosities are converted to $L_{\rm bol, AGN}$ using the bolometric correction $K_{X}$ from \citet{2020A&A...636A..73D}:
\begin{equation}
K_X(L_{2-10\, \rm keV}) \;=\; 15.33 \,\left[\,1 \;+\; 
\left(\frac{\log \!\bigl(L_{2-10\, \rm keV}/L_{\odot}\bigr)}{11.8}\right)^{16.2}\right]
\label{eq:bolcorr}
\end{equation}

\subsection{Emission line contamination}

In \citet{Haidar+24}, we discussed in detail how JWST/MIRI broadband imaging can be subject to contamination from strong emission lines, and as a result, these images may not be exclusively tracing dust emission. For example, the F1000W filter is mainly affected by contributions from the [\ion{S}{IV}]$\lambda 10.51\,\rm \mu m$ and $H_{2}$ 0-0 S(2) [$9.65\, \rm \mu m$]  lines.
This can make it challenging to determine the morphology and extent of the dust, potentially attributing features to dust that may, in fact, be closely linked to ionised or molecular gas instead. \\

We developed a method to correct for contamination, using ESO 428$-$G14 as a case study \citep[see][]{Haidar+24}, successfully producing a line-free image in the F1000W filter. The decontamination method was further extended by \citet{Campbell+25}  to include the full Cycle 1 JWST/GATOS sample and all MIRI filter bands in the ``Dust in the Wind'' programme. \citet{Campbell+25} leverage JWST/MRS spectroscopy available for three of the galaxies (NGC 5728, NGC 3081, NGC 7172) to reproduce synthetic broadband images and cross-match them with the MIRI imaging. By using this technique, they are able to accurately separate the dust continuum from the emission lines and correct the real images accordingly. In the absence of JWST/MRS,  \citet{Campbell+25} find that the imaging-based only method presented in \citet{Haidar+24} systematically overestimates the emission-line contribution by $5-10\, \%$, making it a conservative but reliable measure for dust emission. The images presented in this paper are corrected using this improved method.  We note that, while MRS spectroscopy provides an excellent tool to constrain the contribution of emission lines to the MIRI filters, the spatial resolution is typically $\sim 25\%$ worse than the imager, making  the latter more suitable for resolving fine dust structures  \citep[][]{Campbell+25}. For a full treatment of the contamination correction method and its effects on our images, please see \citet{Campbell+25}.  \\

For morphological studies, we exclusively use images in the F1000W filter, as this band offers the best compromise between spatial resolution with JWST and the prominence of the nuclear extended MIR emission. Longer wavelength images have progressively worse resolution from 15-21 $\rm \mu$m, yielding lower detail in the extended emission. The F560W images offer the highest resolution, but our previous studies show that dust around the AGN does not emit strongly in this band \citep{Haidar+24,Campbell+25}.  \\

\begin{figure*}
  \centering
  {\includegraphics[width=0.95\textwidth]{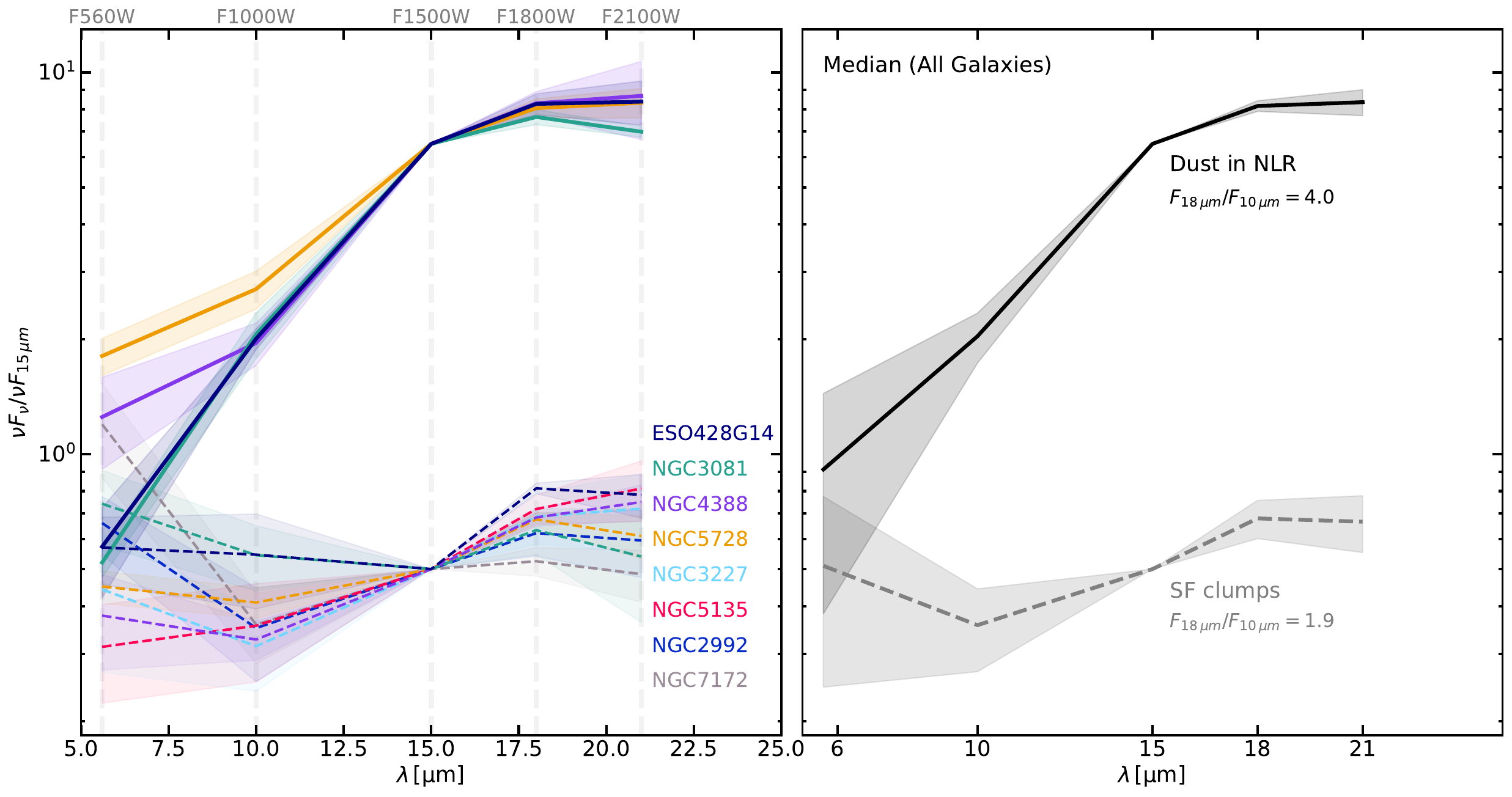}}
   \caption{\textit{Left:} Spectral energy distributions taken over five filters (F560W, F1000W, F1500W, F1800W, F2100W) normalised by the F1500W filter.  Different colours represent different galaxies.  Each curve shows the median SED of all ROIs for that galaxy. For each galaxy, the dashed lines represent the median SED of star forming regions. Where relevant, we plot the median for regions covering dust in the NLR, represented as solid lines. The latter have been shifted up in the Y-axis for clarity. The shaded area represents the 16-84 percentile. \textit{Right:} Galaxy median SEDs for dust in the NLR (solid) and for star-forming clumps (dashed). }
  \label{Median_SEDs_allgals_polar_vs_SF}
\end{figure*}

\section{Results}
\label{results}
\subsection{DUST IN THE NARROW LINE REGION}
\label{sec:dustyNLRs}
We present the full JWST/MIRI imaging sample in Figs.~\ref{dust_in_NLR_AGN} and ~\ref{NO_dust_in_NLR_AGN}.
The images correspond to the F1000W continuum  (dust only) emission, following PSF subtraction and correction for emission line contamination \citep{Haidar+24, Campbell+25}. 
In each galaxy, the MIR emission reveals a range of distinct features, including star-forming disks, either seen face-on (e.g. NGC 5135, NGC 5728) or edge-on (e.g. NGC 7172, NGC 2992), as well as nuclear rings (e.g. NGC 3081, NGC 3227). Each galaxy displays a distinct morphology in the MIR with a wealth of unique features. At the same time, they are all characterised by the presence of a bright nuclear component that, in some cases, outshines the entire galaxy disk. This nuclear component may include emission from the unresolved torus region and will be investigated in detail in follow-up studies focused on individual galaxies.
In what follows, we focus on the central $4''\times 4''$ region ($\sim 300-800\, \rm pc$, depending on the galaxy), which covers the circumnuclear region and matches SINFONI's field of view that maps the ionised [\ion{Si}{VI}] emission. \\

Below, we describe each galaxy in turn, highlighting whether the dust morphology aligns with the NLR axis as traced by ionised and molecular gas. Here, we define \textit{coupling} as the spatial and morphological correspondence between dust and ionised and/or molecular outflows, implying that the dust may be entrained or otherwise influenced by the NLR. \\

\subsubsection{Strong dust–NLR coupling}

Starting with ESO 428$-$G14 (Fig.~\ref{dust_in_NLR_AGN}, top-left panel), the dust emission reveals an asymmetric morphology, extending along a position angle (PA, measured North to East) of $\sim$131$^{\circ}$ from the NW to the SE, with a total extent of  $\sim$250 pc. This emission follows  closely the direction of the radio jet \citep{1996ApJ...470L..31F,1998ApJ...502..199F} and known molecular and ionised  outflows in this AGN \citep[e.g.][]{2018MNRAS.481L.105M,2020ApJ...890...29F}. This indicates that dust is spatially coupled with the ionised NLR gas and the radio jet, as also demonstrated in \citet{Haidar+24}. \\

NGC 4388 (Fig.~\ref{dust_in_NLR_AGN}, top-right panel) is another complex system, featuring dust structures that extend from NE to SW along a PA of $\sim$30$^{\circ}$. The bulk of the extended dust emission is concentrated SW of the nucleus, reaching distances up to $\sim$150 pc. Towards the NE, the dust structure splits into two thin and faint strands, with the longer one extending up to $\sim$150 pc and the shorter one to $\sim$90 pc. Recent mid- to far-infrared imaging with SOFIA further supports this picture, revealing extended $30$–$40\, \mu$m dust emission coincident with the NLR at a similar PA \citep{2025ApJS..276...64F}. This orientation is also consistent with that of the NLR and the radio jet \citep[e.g.][]{2017MNRAS.465..906R}, indicating that the dust is well coupled with the NLR.  \\

In NGC 3081 (Fig.~\ref{dust_in_NLR_AGN}, bottom-left panel), the morphology of the dust traces an S-like shape extending North to South, and along a PA similar to that of the NLR (PA $\sim$165$^{\circ}$).  To the NE of the AGN, the dust emission peaks prominently at $\sim$100 pc, 
coinciding with the region where \citet{2000ApJS..128..139F} previously reported an enhancement in the [\ion{O}{III}]/H$\alpha$ ratio with values as bright as the nucleus itself. The dust overlaps with the ionised bipolar outflow detected by \citet{2016MNRAS.457..972S} within the inner $\sim$200 pc region, further supporting that the dust is coupled with the NLR. \\

Similarly, NGC 5728 (Fig.~\ref{dust_in_NLR_AGN}, bottom-right panel) also exhibits a symmetric biconal dust structure. The latter extends up to $\sim$150 pc from the nucleus on both sides of the cone, oriented at a PA of $\sim$120$^{\circ}$. The dust emission enhances the edges of the cones, taking on a shape that traces the molecular and ionised outflows propagating through the disk \citep{2019MNRAS.490.5860S,2024A&A...689A.263D,2024A&A...691A.162G}. To the NW, a prominent slab-like feature at $\sim$330 pc appears facing the northern cone, also detected in the neon lines with JWST/MRS \citep[][their Fig. 3]{2024A&A...689A.263D}. This further supports the case in which the dust is physically coupled with the ionised NLR in this galaxy.  \\

\subsubsection{Weak or unclear dust–NLR coupling}
In contrast, the remaining galaxies show bright MIR nuclei and/or rings, but with no clear evidence that dust is linked to the NLR and/or outflows. 
In the case of NGC 7172 (Fig.~\ref{NO_dust_in_NLR_AGN}, top-left panel), the MIR emission is dominated by  the prominent edge-on dust lane. Within the inner $1''\times 1''$, a faint hint of extended dust can be seen SW to the nucleus, but it is embedded in the PSF and detectable only after subtraction. While previous studies have established the presence of ionised cones and outflows along the NLR in this galaxy \citep{2017ApJS..232...11T,Alonso-Herrero+23,2024A&A...690A.350H}, the nearly face-on orientation of the ionised cone, combined with the edge-on disk, makes NGC 7172 a case of weak coupling \citep{2024A&A...690A.350H}. This further supports that the dust within the  inner few hundred parsecs in this galaxy may not be well coupled with the NLR. \\

NGC 2992 (Fig.~\ref{NO_dust_in_NLR_AGN}, top-right panel) shows a similar behaviour to NGC 7172, where the MIR emission is dominated by the disk \citep{2015MNRAS.449.1309G}. While the inner $1''\times 1''$ shows a bright nuclear source, there is no evidence of extended dust along the NLR (PA $\sim$125$^{\circ}$).  Interestingly, as shown in Fig.~\ref{NGC2992_ALMA_JWST}, we detect very faint dust clumps along this PA at kpc scale distances from the nucleus, visible only after strong contrast stretching, which coincide with the outflowing CO clumps reported \citep[][their Fig. 6]{2023A&A...679A..88Z}. This is in a region known to have powerful [\ion{O}{III}] outflows \citep{2021MNRAS.502.3618G,2023A&A...679A..88Z}, providing tentative evidence that some dust may be entrained on kpc scales. In the central $\sim$200 pc, the dust emission comes mainly from the disk and the [\ion{Si}{VI}] emission appears compact, with no clear evidence that dust and the NLR are connected at these scales. \\

Another special case is NGC 3227 (Fig.~\ref{NO_dust_in_NLR_AGN}, bottom-left panel), which is nearly face-on and hosts a circumnuclear stellar ring  \citep{2006MNRAS.371..170B}. The ring dominates the MIR emission with PA of $\sim - 30^{\circ}$, as also detected in the near-infrared \citep{2006ApJ...646..754D}. The NLR is defined by an ionization cone oriented along a PA of $\sim$30$^{\circ}$, with evidence of ionised and molecular outflows \citep{2013ApJS..209....1F,Alonso-Herrero+19,2024ApJ...975..129P}. The cone is visible only on the NE side in previous studies \citep{2024ApJ...975..129P}, but it is not  traced in our [\ion{Si}{VI}] map (Fig.~\ref{NO_dust_in_NLR_AGN}), where no distinct ionisation structure is apparent. The cone is almost perpendicular to the nuclear ring, indicating that the dust detected in JWST is not participating in the outflow. \\

Finally, NGC 5135 (Fig.~\ref{NO_dust_in_NLR_AGN}, bottom-right panel) is a strongly barred spiral galaxy, with a central morphology resembling the nuclear ring in HST imaging \citep{2024ApJ...975..129P}. Although the galaxy hosts an extended ionised NLR and outflows \citep{2009ApJ...698.1852B}, the MIR emission arises primarily from the ring with no spatial evidence of it being associated or coupled with the NLR. 
This is in agreement with \citet{2018MNRAS.476.5417S}, where they report that the molecular gas kinematics are dominated by a barred, star-forming structure. \\

This diversity in the sample illustrates that the NLR is a multi-phase medium. All eight galaxies in this programme have been selected based on prior evidence of extended MIR emission on scales of tens to hundreds of parsecs from ground-based observations \citep[][]{2016ApJ...822..109A}. However, such extended emission is not clearly detected in all cases. In ESO 428$-$G14, NGC 4388, NGC 3081, and NGC 5728, the morphology of the dust appears to be closely coupled with the ionised NLR, possibly being entrained in the outflow. On the other hand, NGC 7172, NGC 2992, NGC 3227, and NGC 5153 show weak or inconclusive coupling between the dust and the ionised and/or molecular phases.
It is possible that any dust associated with the NLR or outflow in these galaxies lies on smaller physical scales ($\leq$ 50 pc) and therefore remains unresolved by JWST/MIRI. In general, these galaxies also do not show PAs that favour strong interactions between the NLR and the disk (see Table~\ref{tab:agn_properties}), which would otherwise allow the outflow to easily carry material off the disk. \\

For the remainder of the paper, we focus on the first four galaxies (ESO 428$-$G14, NGC 4388, NGC 3081, and NGC 5728) because they provide the clearest morphological evidence for dust-NLR coupling out to scales of a few hundred parsecs from the nucleus. This spatial correlation between dust, ionised and molecular gas suggests that the dust may be dynamically linked to the NLR, allowing us to investigate its physical properties, survival, and role in AGN feedback.

\begin{figure*}
  \centering
  \includegraphics[width=0.6\textwidth]{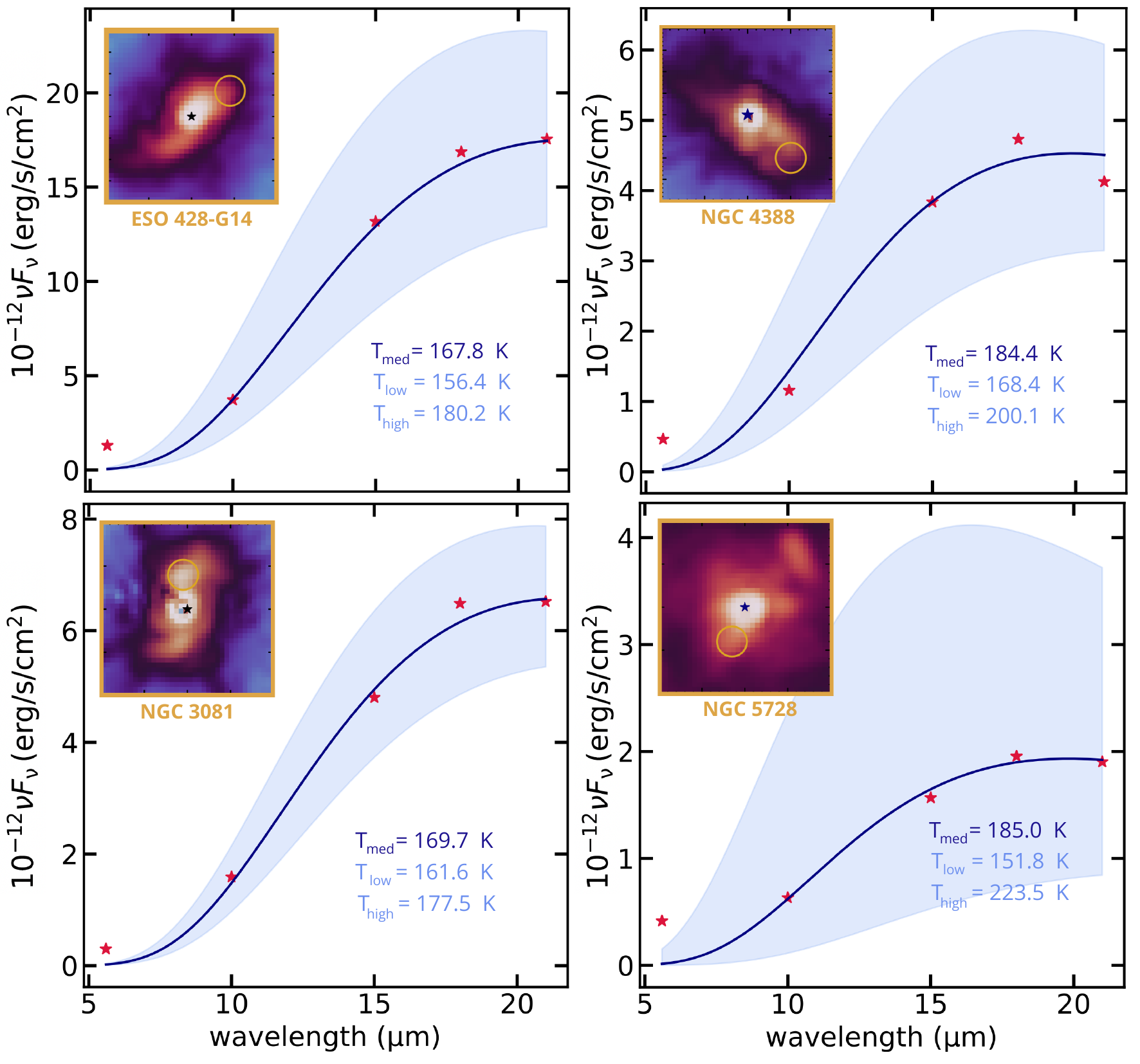}
  \caption{Monte Carlo blackbody fits for the dusty NLRs of ESO~428-G14, NGC~4388 (top row),  NGC~3081, and NGC~5728 (bottom row). Each panel shows an SED extracted with an aperture diameter of $0\farcs7$ at $\sim$150 pc from the nucleus, where shocks dominate (see Fig. ~\ref{dust_in_NLR_AGN}). Apertures are marked as orange circles in the upper left of each panel. Red stars are the photometric points. The solid lines represent the median of the fits and the shaded bands show the 16–84th percentiles. For each galaxy, we show the median temperature $T_{\rm med}$, and the 16–84th values (\(T_{\rm low}\), \(T_{\rm high}\)).}
  \label{MC_DustTemp}
\end{figure*}

\subsection{MIR REGIONS OF INTEREST}
\label{sec:mirclumps}

To assess whether the dust in the NLR is characterised by properties different from other MIR emitting regions, we define several regions of interest (ROIs) across the full sample. From these ROIs we compute photometric SEDs for each galaxy, as shown  in Figs. ~\ref{SEDs_NGC5728}$-$~\ref{SEDs_ESO428}. All apertures are chosen with diameters larger than the PSF FWHM of the longest-wavelength filter (F2100W, FWHM = $0\farcs67$) to minimise the need for wavelength-dependent aperture corrections \citep[see also][]{Haidar+24}. For all ROIs, we adopt an aperture diameter of $0\farcs7$, corresponding to one FWHM of the lowest resolution dataset (i.e. F2100W). We note that going through each ROI and SED galaxy-by-galaxy is beyond the scope of this paper, and we defer this analysis to future work. \\

Fig.~\ref{Median_SEDs_allgals_polar_vs_SF} (left) shows the median SEDs across all ROIs and galaxies, separating dusty NLRs from star forming (SF) clumps. As shown in Section ~\ref{sec:dustyNLRs}, only four galaxies have resolved dust along the NLR (ESO\, 428$-$G14, NGC\, 4388, NGC\, 5728, NGC\, 3081). As a result, we restrict the NLR SEDs to these targets. 
Across all galaxies, we find a systematic difference between SEDs of dust in the NLR and those of SF clumps. On average, SF clumps are characterised by flatter and hotter SED distributions, while dusty NLRs produce SEDs that are steeper by a factor of two (e.g. $F_{18\, \rm \mu m, \, \rm NLR}/F_{10\, \rm \mu m,\,  \rm NLR}$= 4). 
While each ROI in each galaxy shows its own unique distribution, some common trends can still be identified across the sample.
For example, SF clumps consistently produce higher 5.6 $\rm \mu m$ contributions  (e.g. NGC 7172, NGC 3227, NGC 2992). This rise towards the shortest wavelengths may correspond to a low level of stellar light in the MIR. For dusty NLRs, a slight  flux enhancement at 5.6 $\rm \mu m$  is likely due to this stellar component ($\leq 10\%$) \textcolor{blue}{(Rosario+ in prep)}. In general, both NLR dust and SF clumps produce SEDs that peak at around at 18-21 $\rm \mu m$. In SF clumps, a modest excess in the F1800W band over an otherwise flat SED shape could instead arise from Polycyclic Aromatic Hydrocarbon (PAH) emission \citep{2022A&A...666L...5G}.

Overall, we find  the dust in the NLR region emits strongly in the MIR and is physically distinct from SF clumps. This suggests that these two regions may be subject to different radiation fields and/or may host distinct dust grain populations.

\subsection{TEMPERATURE OF DUSTY NLRs}
\label{sec:tdust}

We have shown that the NLRs of ESO 428$-$G14, NGC 4388, NGC 3081, and  NGC5728  are characterised by dusty, distinct morphologies that emit strongly in the MIR. In this section, we quantify this emission by constraining the dust temperature along their NLRs. To do this, we extract SEDs from ROIs located within the NLRs at a common distance from the nucleus. A fixed distance allows us to quantify if AGN properties have an impact on the observed dust temperature. 
We adopt a distance of $d_{\rm AGN} = 150\,\rm pc$ from the nucleus. This distance is large enough such that, in the more distant galaxies, it exceeds the F2100W PSF FWHM ($0\farcs65$), yet not so large such that, in the nearer systems, we do not risk missing regions where NLR dust is still detected.
All apertures are taken with a diameter d=$0\farcs7$, as also done in Section~\ref{sec:mirclumps}.  We note that these regions also overlap with known shock spots from enhanced [\ion{Fe}{II}] emission (see later in Section~\ref{subsec:FeII_emission}). \\

Given we only have five photometric points, it is challenging to constrain the dust temperature from a single black body (BB) fit. As such, we perform a Monte Carlo (MC) fitting approach, where, in each of the 1000 draws, we perturb the fluxes within their uncertainties,  randomly dropping one of the photometric points and fit a single temperature BB to the remaining four. The fits are performed on absolute fluxes, treating the F560W band as an upper limit to account for any potential stellar contribution to the dust emission \textcolor{blue}{(Rosario+ in prep)}. Additionally, 
we adopt  the absolute F560W flux uncertainty as a uniform conservative error across all bands, preventing unrealistically tight temperature constraints given the very small per band errors. 

The solid lines in Fig.~\ref{MC_DustTemp} show the medians of the MC BB fits, with the shaded areas representing the 16–84th percentile distribution. On average, all dusty NLRs produce temperatures within the range of $\sim$$150-200\, \rm K$, consistent with warm dust that peaks in the MIR ($\lambda \sim  \rm 15-19\, \mu m$). The grain population that this represents would depend on the strength of the radiation field heating the dust \citep[e.g][]{Astrodust+23}.\\

Recently, \citet{2025arXiv250719350L} conducted a JWST/MRS continuum study on six nearby galaxies, including NGC 3081, NGC 5728 and NGC 7172. 
They report a relatively constant dust temperature of $\sim$130 K (peaking at 22 $\, \rm \mu m$) within the inner 500 pc, which is lower than the values we derive, albeit measured over a different spatial region. Nevertheless, their study also finds that dust temperatures are broadly consistent across galaxies and show no evident correlation with AGN properties, in agreement with our results.

\begin{figure*}
  \centering
  {\includegraphics[width=0.72\textwidth]{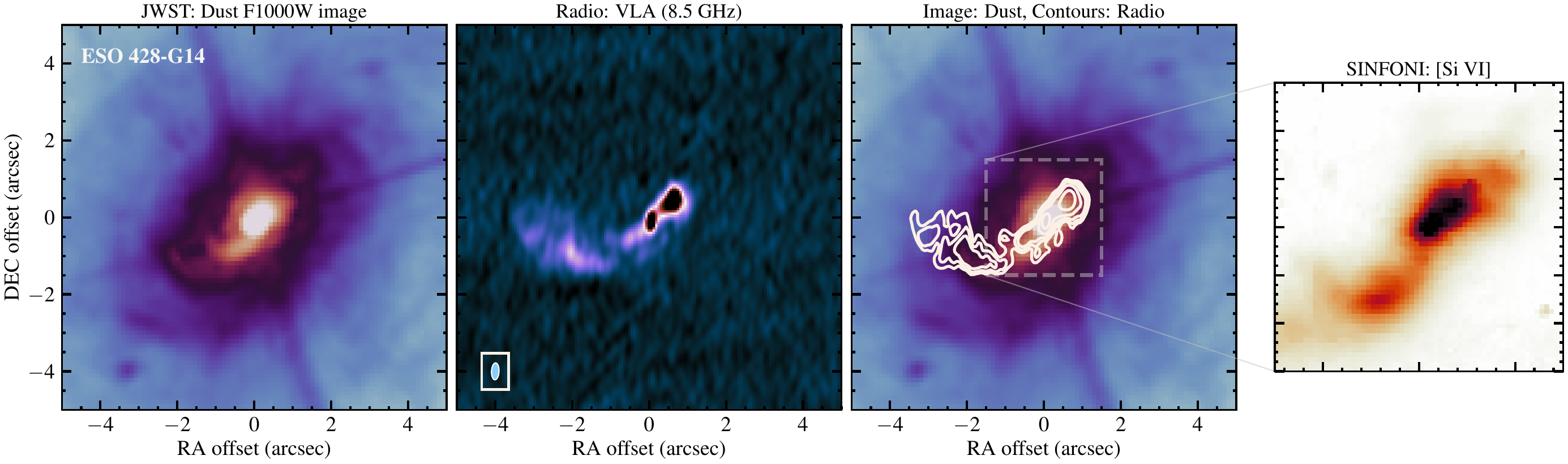}}
  {\includegraphics[width=0.72\textwidth]{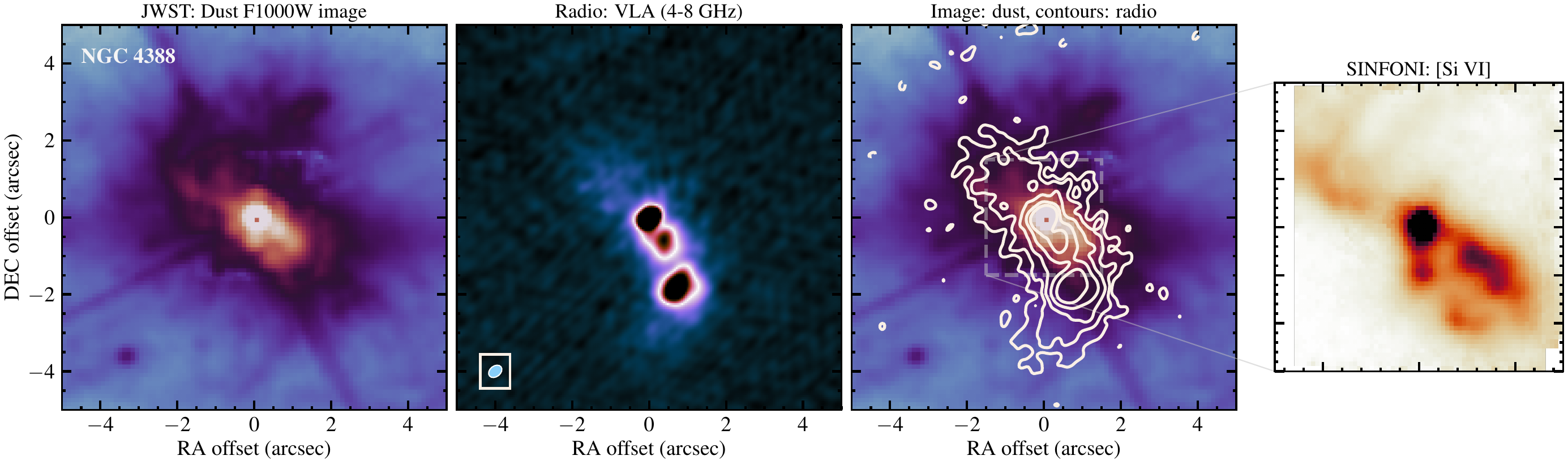}} \\

  {\includegraphics[width=0.72\textwidth]{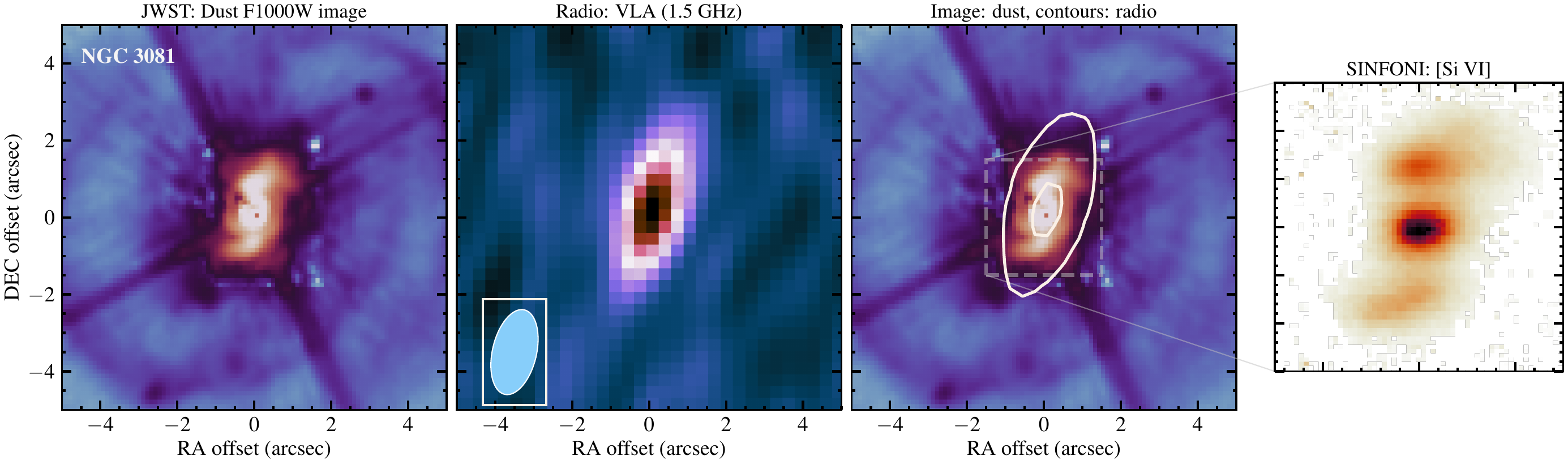}}
  {\includegraphics[width=0.74\textwidth]{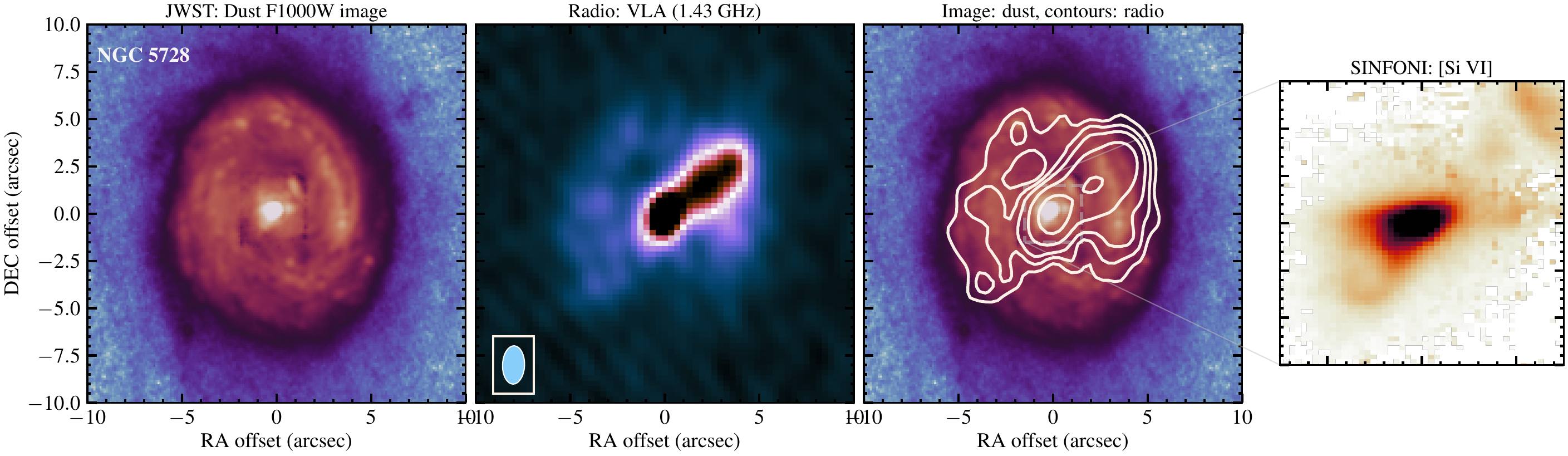}}

  \caption{JWST/MIRI ''Dust in the Wind'' sample. (\textit{Left:}) Extended dust emission revealed by the JWST/MIRI F1000W image, shown after PSF subtraction and decontamination. (\textit{Middle:}) Radio continuum emission from VLA observations. (\textit{Right:}) Comparison between the dust morphology (image) and the radio emission (shown as white contours). The zoomed-in $4'' \times 4''$ region displays  the [\ion{Si}{VI}] coronal line emission from SINFONI.}
  \label{fig:all_radio_dust}
\end{figure*}

\subsection{A DUSTY-RADIO-CORONAL LINK?} 
\label{sec:dustradiolink}

We compare the morphologies of dust, radio, and coronal emission to investigate whether dust near AGN is spatially associated with other AGN-driven components.  In Fig. ~\ref{fig:all_radio_dust}, the images for each galaxy are astrometrically aligned and reprojected onto the largest pixel scale for a common pixel grid. We show the images in their native angular resolutions, rather than convolving to a common PSF, in order to preserve the finer details that each dataset presents.  The different angular resolutions probed by JWST, SINFONI, and the VLA are discussed in detail in Section~\ref{methods}. In most cases, the angular resolution achieved with SINFONI ($\sim0.17''$) is finer than that of JWST's F1000W filter ($\sim0.32''$). For the radio maps where a morphological comparison is relevant (ESO~428–G14: $0.44''\times0.20''$ and   NGC~4388: $0.38''\times0.27''$), the angular resolution is comparable to that of JWST (see Section~\ref{sec:radio-method}). For the purposes of our qualitative morphological inspection, these differences in resolution are not expected to strongly affect our conclusions on the relative extents, nor the orientations, of the emission along the NLR. \\

We find that AGN with dusty NLRs (i.e. ESO\,428$-$G14, NGC 4388, NGC 3081, NGC 5728)  show a strong spatial correlation with the radio emission, as shown in  Fig.~\ref{fig:all_radio_dust}.
Specifically, the radio maps for ESO 428$-$G14 \citep{1996ApJ...470L..31F,1998ApJ...502..199F} and NGC 4388 \citep{2024ApJ...961..230S} exhibit a striking, spatially resolved correspondence with the dust. In both cases, the dust and radio structures extend along the NLR, and their morphologies closely trace one another, as shown in the first two panels of  Fig.~\ref{fig:all_radio_dust}. \\

For NGC 5728, the $1.5\, \rm GHz$ radio emission is prominent along a PA consistent with the NLR and the dust emission, but extends primarily towards the NW \citep{1993ApJ...419L..61W}, reaching several hundred of parsecs beyond the observed dusty bicone. A fainter contribution from the disk is also present, though less prominent than the jet/NLR emission.  NGC 3081, on the other hand, shows only a tentative hint of the $1.5\, \rm GHz$ radio emission along the NLR, but the radio map is not well resolved and is aligned with the direction of the beam ($2.25''\times1.14''$) as also demonstrated in  \citet[][]{1999ApJS..120..209N} and \citet[][]{2009ApJ...703..802M}. As such, we do not draw any conclusions from the map presented in Fig.~\ref{fig:all_radio_dust}. Very recently, \citet{2025ApJ...986..194S} present new VLA 6 GHz imaging of NGC 3081, revealing spatially resolved radio emission that extends well beyond the beam and up to $\sim$170 pc from the nucleus on both sides (see their Fig. 1). The emission reported in \citet{2025ApJ...986..194S} is extended along a similar PA to the dust probed in our study, further supporting a dust-radio link in NGC 3081.  \\

In Fig.~\ref{fig:dust_radio_no_coupling}, we show that several galaxies with weak dust–NLR coupling nevertheless present interesting large-scale radio features. In particular, NGC 2992 exhibits a striking  ``figure-eight'' in the 4.8 GHz radio map. This has been interpreted in the literature as radio bubbles, likely linked to stellar activity \citep{2000MNRAS.314..263C}. Overall, we find no spatial correspondence between the radio bubbles and the dust emission. 
NGC 5135 is another unique case, hosting a radio hotspot  at $\sim$1 kpc from the nucleus, associated with a supernova shock \citep{2009ApJ...698.1852B,2012ApJ...749..116C} which is evident in the 4-8GHz map as a large radio hotspot. While the radio emission traces nicely the dust morphology at larger scales, within the central $4''\times4''$  the galaxy shows no clear link between the dust and the radio along the NLR.  Overall, this work motivates future homogenous radio observations for these key targets. \\

Another compelling connection is that between the dust and the coronal line emission. In Fig.~\ref{fig:all_radio_dust}, we systematically find that the morphology of the dust closely mirrors that of the [\ion{Si}{VI}] emission, regardless of whether there is a tight dust-radio correlation or not.  This is also true for galaxies with weak dust–NLR coupling, as shown in Fig.~\ref{fig:dust_radio_no_coupling}, where both the dust and [\ion{Si}{VI}] emissions appear compact within the inner $4''\times4''$ region. \\

\Needspace{6\baselineskip}
Previous studies have shown that AGN photoionization is generally responsible for the observed coronal line emission in Seyferts. These are highly ionised species,  with high ionisation potentials (IPs > 100 eV), which makes them insensitive to star formation and can therefore act as direct tracers of AGN activity \citep[e.g.][]{2005MNRAS.364L..28P,2006ApJ...653.1098R}. However, when these coronal lines are observed out to several hundreds of parsecs from the nucleus, as in ESO 428$-$G14, NGC 4388, NGC 3081, NGC 5728, AGN photoionization alone is unlikely to account for this. This is because the high-energy photon flux becomes too low at large distances.
Therefore, the detection of extended coronal lines can be used as a tracer for an additional in-situ excitation process, such as shocks \citep[e.g.][]{1997ApJS..110..287F}. \\

More recently, \citet{2025MNRAS.tmp..423R} investigated the extent of the coronal emission in ESO 428$-$G14, NGC 5728, NGC 3081, among other Seyferts, and found a strong positive trend between the size of the coronal line emission and the power of the radio jet. Their work suggests that radio jets propagating through the ambient gas drive shocks that produce coronal lines extending from several hundred parsecs to kiloparsec scales. \\

While it remains debated whether the radio emission in Seyferts traces a jet or a wind \citep[e.g.][]{1996ApJ...470L..31F,1998ApJ...495..680B,2023ApJ...953...87F}, this distinction is not critical here. Regardless of the nature of the radio emission, AGN hosting outflows are capable of driving shocks \citep[e.g.][]{2022A&A...665L..11P,2025ApJ...982...69R,2023MNRAS.525.5575F,2025MNRAS.537.2003F}. 
In addition to the excitation of the coronal line, an important shock tracer is an enhancement in [\ion{Fe}{II}]/Pa$\beta$ and  [\ion{Fe}{II}]/[P II] ratios, which has already been reported in the literature within the central few hundred parsecs of these galaxies. 
We highlight in Fig.~\ref{dust_in_NLR_AGN} the regions where these shocks are likely to be important. In what follows, we summarise some key observational signatures of shocks within the central few hundred parsecs of our main AGN:  ESO 428$-$G14, NGC 4388, NGC 3081 and NGC 5728.
\subsubsection{Evidence from [\ion{Fe}{II}] emission}
\label{subsec:FeII_emission}

The enhancement in [\ion{Fe}{II}]$1.644 \micron$ emission relative to a non refractory species, such as [P II], and evaluated by means of flux ratios [\ion{Fe}{II}]/[P II] or [\ion{Fe}{II}]/Pa$\beta$, offer a key diagnostic to measure the relative contribution of photoionisation and shocks \citep[e.g.][]{2001A&A...369L...5O,2006ApJ...645..148R,2009ApJ...694.1379R,2009MNRAS.394.1148S,2014MNRAS.442..656R}. Maps using these line ratios have also been found to correlate well with radio emission, further supporting that shock excitation may be the key ingredient in the production of [\ion{Fe}{II}] \citep[e.g.][]{1993ApJ...416..150F,1994ApJ...421...92B}.

\begin{itemize}
    \item In ESO 428$-$G14, \cite{2006MNRAS.373....2R} report on the detection of  $\rm 1.27 \leq [\ion{Fe}{II}]/Pa \beta \leq 2.73 $ ratios that are co-spatial with the radio emission. The strongest [\ion{Fe}{II}]/Pa$\beta$ values coincide with the radio hot spots towards the NW, approximately $\sim 120 \, \rm pc$ away from the nucleus.
    \item In NGC 4388, [\ion{Fe}{II}]/Pa$\beta$ peaks SW to the nucleus at $\sim 150\, \rm pc$ \citep{2001AJ....122..764K,2017MNRAS.465..906R}. In \citet{2017MNRAS.465..906R}, they also report on the detection of double peak [\ion{Fe}{II}] lines in the same region, further supporting an additional excitation source other than photoionisation to be at play. 
    \item In NGC 5728, \citet{2018ApJ...867..149D} use [\ion{Fe}{II}]/[P II] to constrain the ionisation source in the NLR. Ratios with values higher than 2 would indicate shock contribution. They compute [\ion{Fe}{II}]/[P II] along the cone and find ratios of $\sim  5.8$ SW to the AGN and $\sim 3.9$ in the NW.  
    \item In NGC 3081, \cite{2003MNRAS.343..192R} leverage the [\ion{Fe}{II}]/Br$\gamma$ ratio, which can also be sensitive to the excitation mechanism, and report a weak value of 0.6. However, the coronal line emission in NGC 3081 appears more spatially complex. \citet{2005MNRAS.364L..28P} resolved the  [Si VII]2.48 $\micron$ emission into a compact source associated with the AGN, and a fainter blob located  $\sim$120 pc NW to the AGN. This same blob coincides with the second brightest region in H$\alpha$  and [\ion{O}{III}], previously reported by \citet{2000ApJS..128..139F} (see their Fig. 11). \citet{2005MNRAS.364L..28P} propose that the concentric shell-like structures seen in H$\alpha$ and [\ion{O}{III}] could be indicative of propagating shock fronts. We mark this region as a tentative shock site in Fig. ~\ref{dust_in_NLR_AGN}.
    
\end{itemize}

Taken together, the close correspondence between dust, radio, and coronal line structures, along with evidence of strong enhancement in [\ion{Fe}{II}] emission, supports the scenario in which shocks play a key role in shaping the NLRs of these AGN. However, the extent to which shocks contribute to the infrared emission, as opposed to direct AGN illumination, remains unclear. Below we investigate the origin of dust heating in the NLR, now that the broader picture of its morphology and excitation have been established.

\section{Discussion}
\label{sec:discussion}

\begin{table}
\centering
\caption{Expected AGN radiation field $U_{\rm AGN}$ and dust temperatures (assuming  classical ISM grain sizes) derived at a distance $d_{\rm AGN} = 150\,\mathrm{pc}$ from the nucleus. }
\begin{tabular}{c c c c }
\hline
AGN & $\log_{10} U_{\rm AGN}$ &  $T_{\mathrm{dust,a=0.005\,\mu m}}$ & $T_{\mathrm{dust, a=0.25\,\mu m}}$ \\ 
 & &  (K) & (K) \\ 
\hline
ES0428-G14 & 3.12  & 64 & 29 \\ 
NGC 4388   & 4.08  &  100 & 46 \\ 
NGC 5728   & 4.38  &  115&  53\\ 
NGC 3081   & 4.36   & 114 & 52\\  \hline
NGC 7172   & 4.24   & 108 & 49\\ 
NGC 2992   & 3.50   & 77 & 35\\ 
NGC 3227   & 3.57   & 80 & 36\\  
NGC 5135   &  4.84  & 142 & 65 \\ 
\hline
\end{tabular}
\label{tab:radfield-agn}
\end{table}

\subsection{WHAT HEATS THE DUST IN THE MIR?}
\label{sec:what-heats-dust}

Given the direct influence of the central source on the NLR, the simplest and most logical interpretation would be  that dust in the NLR is heated and illuminated by AGN radiation fields, as previously reported in the literature for a number of Seyferts. The challenge here, however, is not whether photons are energetic enough, but whether the AGN radiation fields can sustain dust temperatures in the range of $150-200\, \rm K$ at distances of several hundred parsecs from the nucleus. To determine this, we constrain the expected AGN radiation field for these galaxies and assess whether it alone is able to reproduce the  dust temperatures presented in Section~\ref{sec:tdust}. \\

\begin{figure}
  \centering
  {\includegraphics[width=0.5\textwidth]{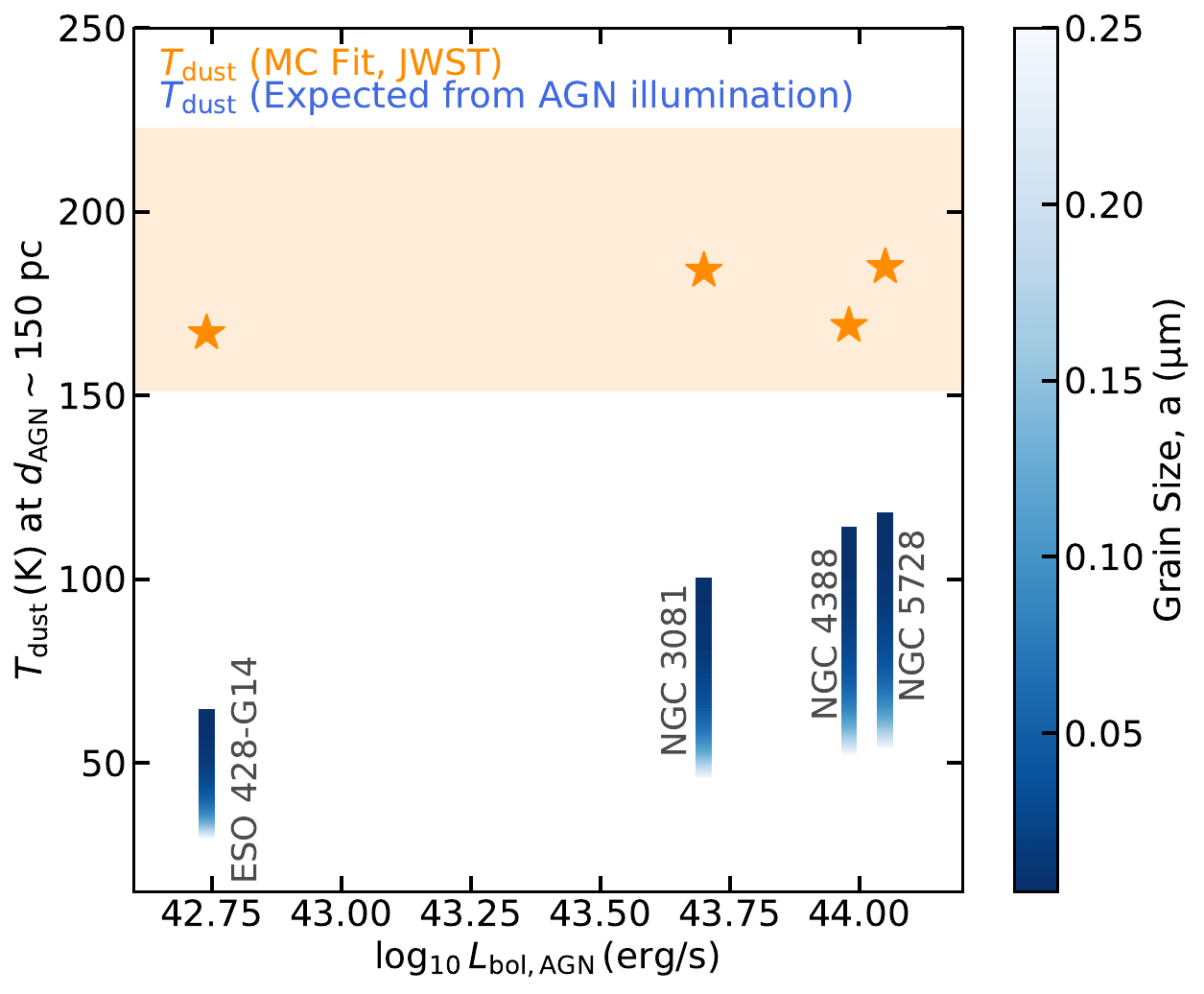}}
  \caption{Comparison between the median dust temperatures from MC BB fits (orange stars) and predicted temperatures from AGN illumination (blue bars). The shaded orange area represents the minimum and maximum possible dust temperatures across the four galaxies. The colour gradient within each blue bar represents the range of predicted temperatures resulting from variations in the assumed grain size $a = [0.005,0.25]\, \micron$. For all four galaxies, additional heating mechanisms must be present, as AGN illumination alone cannot account for the observed dust temperatures.}
  \label{dusttemp_AGNillumination}
\end{figure}

Following \citet{Tielens+10} (Chapter 5, Eq. 5.43), and as also done in \cite{Haidar+24}, we calculate the AGN radiation field at a distance $d_{\rm AGN}$ from the nucleus in terms of the expected anisotropic radiation field expressed in units of the Habing field: 
\begin{equation}
    U_{\rm AGN} = 2.1 \times 10^4 \left( \frac{L_{\rm bol, AGN}}{10^4 L_{\odot}} \right) \left( \frac{0.1\, \text{pc}}{d_{\rm AGN}} \right)^2,
\label{eq:U_AGN}
\end{equation}
where $L_{\rm bol, AGN}$ is the bolometric AGN luminosity (see Section~\ref{sec:bol}).  Subsequently, for a given grain size $a$, the dust temperature can be expressed as:
\begin{equation}
    T_{\rm d} \approx 33.5 \left( \frac{1\, \mu m}{a} \right)^{0.2} \left( \frac{U_{\rm AGN}}{10^4} \right)^{0.2} \, \rm K,
\label{eq:tdust}
\end{equation}
\noindent following \citet{Tielens+10} (Chapter 5, Eq.~5.44). Note that this assumes that the intrinsic AGN SED that heats the dust peaks strongly in the UV, roughly like the local Habing field, which corresponds to $1.6\times10^{-3}\, \rm erg \, cm^{-2} s^{-1}$ \citep{1968BAN....19..421H}. We note that AGN can also show an enhanced soft X-ray component, however this would be inefficient at heating dust, as grain absorption efficiencies drop steeply at X-ray energies \citep[e.g.][]{1993ApJ...402..441L}.

We compute the AGN radiation field at a fixed distance of $d_{\rm AGN} = 150 \, \rm pc$, the same as that adopted in Section~\ref{sec:tdust}. This allows us to constrain whether  the temperatures expected from AGN heating are able to reproduce those derived from the observations.  
We present the results for the full sample in Table~\ref{tab:radfield-agn}. On average, we find that the AGN radiation field across the sample to be $\log_{10} U_{\rm AGN} \sim 4$. This corresponds to dust temperatures of $T_{\rm dust} \sim 100\,\rm  K$ for the smallest grains (i.e, $a = 0.005\, \micron$) and $T_{\rm dust} \sim 50\, \rm K$ for larger grains (i.e, $a = 0.25\, \micron$), consistent with the canonical grain size distribution proposed by \citet{1977ApJ...217..425M}. \\

We summarise our results in Fig.~\ref{dusttemp_AGNillumination}. The orange stars represent the dust temperatures computed from the SEDs of the shocked regions indicated in Fig.~\ref{MC_DustTemp}, and the shaded area covers the lowest and highest possible temperatures across all four galaxies $T_{\rm dust} \sim [150, 220]\, \rm K $. The blue column represents the possible dust temperatures over a range of grain sizes $a = [0.005,0.25]\, \micron$ expected from AGN heating.

In general, for a fixed distance from the nucleus, a higher bolometric luminosity leads to higher dust temperatures. However, even when accounting for the smallest dust grain sizes, which tend to heat more efficiently, the temperature expected from AGN heating still does not reach those probed  with JWST. To achieve a minimum temperature of $T \sim 150\, \rm K$, one would  require AGN luminosities in the range of  $L_{\rm AGN, bol} = [3.8 \times 10^{44}, 2\times 10^{46}]\, \rm erg/s$ for grains  $a\,= [0.005,0.25]\, \micron$, respectively.  Values $\geq 10^{45}\, \rm erg/s $ enter the quasar regime and are atypical of local Seyferts. We thus interpret this as excess heating that cannot be accounted for by AGN illumination alone, indicating that additional in-situ heating may be at play.  Below we discuss possible processes in more detail.

\subsection{IN-SITU DUST HEATING}
\label{sec:insitu_heating}

Several processes can, in principle, contribute to heating the dust in the NLR. Such processes include: in-situ heating from ongoing star formation, radiation pressure driven polar wind, radiative shocks from the post-shock region, or simply, collisional heating from hot plasma. 

We have shown in Fig.~\ref{Median_SEDs_allgals_polar_vs_SF} that dusty NLRs produce SEDs that are systematically steeper than those of SF clumps. As such, we can rule out star formation as a dominant source of radiation in the NLR. Polar dusty winds, while effective on parsec to tens of parsec scales \citep{2019ApJ...884..171H}, cannot propagate to the hundreds of parsecs scale probed here, consistent with interferometric results and radiative transfer models \citep[e.g.][]{2019ApJ...884..171H,Alonso-Herrero2021}. \\

Given the strong spatial correspondence between dust in the NLR and the radio emission (Fig.~\ref{fig:all_radio_dust}), jet and/or wind-ISM interactions, such as shocks, could contribute to the observed MIR emission. In fact, \citet{1981ApJ...245..880D} report that for fast shocks with velocities ($v_{\rm shock} \ge 200\, \rm km/s$), dust grains become an important coolant, radiating up to $17\, \%$ of the shock energy in the infrared. As presented in Section~\ref{sec:dustradiolink},  all AGN with dusty NLRs (ESO 428$-$G14, NGC 4388, NGC 3081, and NGC 5728) have direct evidence of shocks in the central few hundred parsecs (see regions highlighted in Fig.~\ref{dust_in_NLR_AGN}). This further establishes shocks as a strong candidate for additional dust heating. \\

By adopting the \citet{1995ApJ...455..468D} formalism for fast radiative shocks, we demonstrated in \citet{Haidar+24} that the post-shock cooling region can generate radiation fields strong enough to heat the dust to MIR-emitting temperatures. As an exercise, we apply the same approach here to test which combinations of shock velocity and ISM density are needed to reproduce the minimum dust temperatures observed in our sample.
The total radiative flux per unit shock surface is given by: 

\begin{equation}
   F_T = 2.28 \times 10^{-3} \left( \frac{v_s}{100 \, \text{km s}^{-1}} \right)^{3.0} \left( \frac{n}{\text{cm}^{-3}} \right) \, \text{ergs cm}^{-2} \text{s}^{-1},
\end{equation}

\noindent
where $v_s$ is the velocity of the shock and $n$ is the particle number density. 
Substituting this flux into Eq.~\ref{eq:tdust}, scaling it by the Habing field, and solving for  $T_{\rm dust} = 150\, \rm  K $ and a= 0.005 $\micron$, implies $v_{\rm shock}\sim 400 \, \rm km/s$ for $n\sim  10^{3} \rm cm^{-3}$, $v_{\rm shock}\sim 230 \, \rm km/s$ for $n\sim 5 \times 10^{3} \rm cm^{-3}$, and $v_{\rm shock}\sim 180 \, \rm km/s$ for $n\sim  10^{4} \rm cm^{-3}$. These estimates show that, shock velocities in the range of $v_{\rm shock} = 200 - 400 \, \rm km/s$ propagating through dense enough media ($n\sim  10^{3-4} \rm cm^{-3}$), could already account for dust temperatures of $\sim 150\, \rm K$.

Recent JWST/MRS studies compared shock models with MIR line ratio diagnostics for a number of AGN, including NGC 3081 and NGC 5728, and found significant overlap between the predictions of  photoionisation and shock models, making it difficult to set robust constraints on shock velocities and predensities \citep[e.g.,][]{2025ApJS..280...65Z, 2025arXiv251002517R}. 
Nevertheless, such conditions are reasonable and have already been reported for the low-lumosity AGN ESO 428$-$G14 \citep{2006MNRAS.373....2R,2018MNRAS.481L.105M}. This makes shock heating not only a plausible additional source, but also a mechanism that could potentially by itself account for the observed dust temperatures.  Future work will explore shock modelling tailored specifically for JWST/MRS observations. \\

Another interesting process to consider is collisional heating of dust grains by hot plasma. This is especially relevant given the strong morphological resemblance between   the dust and the coronal line emission (see Fig.~\ref{dust_in_NLR_AGN}), which traces hot gas in the range $T \sim 10^{5-6}\, \rm K$ \citep[e.g.][]{1994A&A...288..457O}. In such hot plasma, collisions of dust grains with electrons/ions can drive dust to higher effective temperatures \citep[e.g.][]{1979ApJ...231..438D,1987ApJ...322..812D,2013A&A...556A...6B,2019Ap.....62..540D}. This process however strongly depends on grain size, as smaller grains undergo large temperature fluctuations but are also quickly destroyed by charging and sputtering, whereas larger ones could survive long enough to contribute to the dust emission \citep[][]{1979ApJ...231..438D,1987ApJ...322..812D}.

\subsection{A SURVIVING POPULATION OF DUST GRAINS?}

The presence of dust in regions dominated by shocks (Fig.~\ref{dust_in_NLR_AGN}), as indicated by strong enhancement in [\ion{Fe}{II}] emission (Section~\ref{subsec:FeII_emission}), raises important questions about dust survival under such conditions.
Theoretically, shocks are expected to destroy dust grains, releasing refractory elements such as [\ion{Fe}{II}], previously locked within them, back into the gas phase \citep[][]{1952ApJ...115..227R,1978ApJ...225..887C,1997ApJS..110..287F,1994ApJ...433..797J,1994ApJ...431..321T,1996ApJ...469..740J}.
However, 
in a number of systems, observations show that dust is prevalent where shock destruction processes such as sputtering and/or shattering are expected
\citep[e.g.][]{1990MNRAS.246..163T, 2000A&A...363..926A,2001MNRAS.328..848V,shahbandeh2023jwst,Haidar+24}. The survival of dust in extreme conditions has attracted considerable attention in recent years, with models predicting factors such as longer destruction timescales, grain size distribution and composition, coagulation, as well as shock velocity to play  key roles in supporting dust survival \citep[][]{2014A&A...570A..32B, 2017ApJ...850...40R, 2022MNRAS.510..551F, 2024arXiv240303711R, 2024MNRAS.528.5008O, 2024A&A...687A.240D}.

The survival of dust has  direct implications on the survival of $H_{2}$ molecular  gas. The latter has a dissociation potential of only $\sim$5 eV and would not survive direct AGN irradiation without dust shielding. This has led some models to suggest that molecules may instead form inside the outflows, with dust acting as the necessary catalyst \citep[e.g.][]{2014MNRAS.439..400Z,2018MNRAS.474.3673R,2018MNRAS.478.3100R}. Grain growth is possible in the outflow through metal accretion and/or coagulation, each leaving a specific imprint on the extinction curve \citep[e.g.][]{2012MNRAS.422.1263H,2017P&SS..149...45H}. On the other hand, complete dust formation (i.e. production of grain seeds) in the outflows is much more difficult as it requires gas conditions that are similar to AGB stars or supernova ejecta, where very dense, metal-rich and rapidly cooling gas would enable the nucleation of stable dust precursors \citep[e.g.][]{2002ApJ...567L.107E,2019ApJ...885..126S}. \\

Dust processing also strongly depends on grain material and composition. In particular, sputtering and shattering destruction thresholds are set by material bond properties \citep[][]{1994ApJ...431..321T}. For example, in contrast to carbonaceous grains, silicate grains are thought to be difficult to reform under ISM conditions and are therefore therefore believed to originate in AGB stars. As such, silicate grains should be preserved in the passage of shocks \citep{2011A&A...530A..44J}. In SN shocks, \citet{2014A&A...570A..32B} find that carbonaceous grains are quickly destroyed, even when exposed to slow shocks $v_{\rm shock} \sim 50\, \rm km/s$, while silicate grains demonstrate greater resilience to shock processing.
As such, a possible hypothesis would be that shocks  preferentially destroy carbonaceous grains, leaving a population dominated by silicates in post-shock regions. This picture is further complicated by the fact that sputtering yields are higher for silicates than for graphite \citep[e.g.][]{1994ApJ...431..321T}, which would suggest the opposite trend. However, survival in shocks depends on more than sputtering yields alone, including grain charging, shattering, and the ability of different materials to reform in the ISM. This gap remains poorly explored and highlights the need for more realistic dust modelling in shocked environments to constrain how grains endure the passage of a shock.

\section{Summary \& Conclusion}
\label{conclusion}

We investigated the presence and origin of dust in the narrow line regions of  eight nearby Seyferts from the JWST/MIRI imaging programme \textit{``Dust in the Wind''}, part of the GATOS survey. Our key findings can be summarised as follows:

\begin{itemize}
 
    \item Among the eight galaxies, ESO 428$-$G14, NGC~4388, NGC~3081, and NGC~5728 show extended dust structures within the inner $r \sim 2''$ region, aligned with the narrow line region and exhibiting morphologies comparable to the ionised gas (see Figs.~\ref{dust_in_NLR_AGN} and \ref{NO_dust_in_NLR_AGN}).

    \item For galaxies with dusty narrow line regions, we find a tight spatial correspondence between the morphology of the dust, the radio  and the [\ion{Si}{VI}] emissions (see Fig.~\ref{fig:all_radio_dust}). Indeed, all of these galaxies host narrow line regions oriented at inclinations (see Table~\ref{tab:agn_properties}) that favour coupling between the extended dust, the ionised gas, and the outflows.

    \item We extract photometric SEDs across several regions of interest and find that dust in the narrow line region produces systematically steeper SEDs than star-forming clumps, with $F_{18\, \rm \mu m}/F_{10\, \rm \mu m}$ ratios of $\sim4$ versus $\sim2$ respectively, averaged across the full sample (Fig.~\ref{Median_SEDs_allgals_polar_vs_SF}). This suggests that dust in the narrow line region is subject to different grain processing and/or harder radiation fields than the surrounding regions.

    \item We compute the temperature of the dust in shock-dominated regions within the narrow line region, at 150 pc from the nucleus, and find consistent values within the range of  150–200 K, characteristic of warm dust peaking in the MIR (Fig.~\ref{MC_DustTemp}).  Such temperatures can be easily achieved if the emission is dominated by small grains ($a =~ 0.005\, \rm \mu m$). 

    \item Using simple models, we investigate the origin of the observed dust emission and find, even under optimistic assumptions of grain size and AGN luminosity, direct AGN heating alone is not able to reproduce the observed dust temperature. We propose that the excess heating is instead produced in-situ. We show that fast radiative shocks with velocities of $v \sim 200$–$400 \, \rm km\ s^{-1}$ propagating through dense enough media ($n \sim 10^{3-4}\,\rm cm^{-3}$) could account for the observed temperatures. \\

    The evolution and processing of dust in shocks depends on several parameters, such as grain size, composition, local density, and shock velocity, that are often not well constrained in observations. This highlights the need for shock and dust evolution models to evaluate how efficiently these properties contribute to the observed dust in AGN-driven outflows.  Future progress will come from combining such dust models with constraints from  JWST and ALMA, providing multi-phase diagnostics needed to build a complete picture on dust survival. 
    Already, the tentative detection of faint warm dusty clumps in NGC 2992 (see Fig.~\ref{NGC2992_ALMA_JWST}  and Section~\ref{ap:ngc2992}) that can be cross matched with outflowing CO clumps \citep{2023A&A...679A..88Z} provide a glimpse of how the warm and cold phases interconnect in outflows, a link that would be critical to fully understand dust survival and its role in AGN feedback.

\end{itemize}



\section{DATA AVAILABILITY}
All data is publicly available and can be extracted as described in Section \ref{methods}. The reduced JWST images are available upon request, otherwise can also be extracted from MAST according to Section \ref{methods}.

\section{Acknowlegments}
We thank the referee for a constructive report that helped improve the manuscript.
HH thanks Martin Elvis and Brandon Hensley for valuable discussions on dust processing, Maria Vitoria Zanchettin for providing the NGC 2992 ALMA maps, and Travis Fischer and the \citet{2024ApJ...961..230S} team for sharing the NGC 4388 radio maps. HH is grateful to ESA/ESAC for hosting her research visit in Madrid, with special thanks to  Alvaro, Ismael, Almudena, and Laura for their support during the visit. HH acknowledge support from ESA through the Science Faculty (Funding reference ESA-SCI-E-LE-183), and  funding from STFC and Newcastle University.

\noindent
DJR \& SC acknowledge the support of the UK STFC through grant ST/X001105/1.
AAH, LHM and MVM acknowledge support from grant PID2021-124665NB-I00 funded by the Spanish
Ministry of Science and Innovation and the State Agency of Research
MCIN/AEI/10.13039/501100011033  and ERDF A way of making Europe.
AA also acknowledges funding from the European Union (WIDERA ExGal-Twin, GA 101158446).
MPS acknowledges support under grants RYC2021-033094-I, CNS2023-145506 and PID2023-146667NB-I00 funded by MCIN/AEI/10.13039/501100011033 and the European Union NextGenerationEU/PRTR.
IGB is supported by the Programa Atracci\'on de Talento Investigador ``C\'esar Nombela'' via grant 2023-T1/TEC-29030 funded by the Community of Madrid.
CR acknowledges support from SNSF Consolidator grant F01$-$13252, Fondecyt Regular grant 1230345, ANID BASAL project FB210003 and the China-Chile joint research fund. 
MS acknowledges support by the Ministry of Science, Technological Development and Innovation of the Republic of Serbia (MSTDIRS) through contract no. 451-03-66/2024-03/200002 with the Astronomical Observatory (Belgrade).
CRA and AA acknowledge support from the Agencia Estatal de Investigaci\'on of the Ministerio de Ciencia, Innovaci\'on y Universidades (MCIU/AEI) under the grant ``Tracking active galactic nuclei feedback from parsec to kiloparsec scales'', with reference PID2022$-$141105NB$-$I00 and the European Regional Development Fund (ERDF).
AJB acknowledges funding from the ``FirstGalaxies'' Advanced Grant from the European Research Council (ERC) under the European Union’s Horizon 2020 research and innovation program (Grant agreement No. 789056).
DR acknowledges support from STFC through grants ST/S000488/1 and ST/W000903/1.
C.P., J.S, L.Z., E.K.S.H., acknowledge grant support from the Space Telescope Science Institute (ID: JWST-GO-02064.002).
FE and SGB acknowledge support from the Spanish grant PID2022-138560NB-I00,
funded by MCIN/AEI/10.13039/501100011033/FEDER, EU. 
C.M.H acknowledges funding from a United Kingdom Research and Innovation grant (code: MR/V022830/1).
OG-M acknowledge financial support from PAPIIT UNAM project IN109123 and “Ciencia de Frontera” CONAHCyT project CF-2023-G-100. 
ARA acknowledges Conselho Nacional de Desenvolvimento Científico e Tecnológico (CNPq) for partial support to this work through grant 313739/2023-4.
E.B.acknowledges support from the Spanish grants PID2022-138621NB-I00 and PID2021-123417OB-I00, funded by MCIN/AEI/10.13039/501100011033/FEDER, EU.
MJW acknowledges support from a Leverhulme Emeritus Fellowship,
EM-2021-064. 
RAR  acknowledges the support from the Conselho Nacional de Desenvolvimento Científico e Tecnológico (CNPq; Projects 303450/2022-3,  and 403398/2023-1),  the Coordenação de Aperfeiçoamento de Pessoal de Nível Superior (CAPES; Project 88887.894973/2023-00), and Fundação de Amparo à Pesquisa do Estado do Rio Grande do Sul (FAPERGS). 
SFH acknowledges support through UK Research and Innovation (UKRI) under the UK government’s Horizon Europe Funding Guarantee (EP/Z533920/1, selected in the 2023 ERC Advanced Grant round) and an STFC Small Award (ST/Y001656/1).

This work is based on observations made with the NASA/ESA/CSA James Webb Space Telescope.
This work makes use of several \texttt{PYTHON} packages:  \texttt{NumPy} \citep{Harris2020}, \texttt{Matplotlib} \citep{Hunter2007}, and \texttt{Astropy} \citep{Astropy2013,Astropy2018}. \\

\bibliographystyle{mnras}
\bibliography{references} 

\hfill \\

\noindent
$^{1}$ School of Mathematics, Statistics and Physics, Newcastle University, Newcastle upon Tyne, NE1 7RU, UK\\
$^{2}$ Centro de Astrobiolog\'{\i}a (CAB), CSIC--INTA, Camino Bajo del Castillo s/n, E\mbox{-}28692 Villanueva de la Ca\~nada, Madrid, Spain\\
$^{3}$ Instituto de Astrof\'{\i}sica de Canarias, Calle V\'{\i}a L\'actea s/n, E\mbox{-}38205 La Laguna, Tenerife, Spain\\
$^{4}$ Departamento de Astrof\'{\i}sica, Universidad de La Laguna, E\mbox{-}38206 La Laguna, Tenerife, Spain\\
$^{5}$ LUX, Observatoire de Paris, Coll\`ege de France, PSL University, CNRS, Sorbonne University, Paris, France\\
$^{6}$ Department of Physics, University of Oxford, Keble Road, Oxford, OX1 3RH, UK\\
$^{7}$ Department of Astronomy, University of Geneva, ch. d'Ecogia 16, 1290, Versoix, Switzerland\\
$^{8}$ Instituto de Estudios Astrof\'isicos, Facultad de Ingenier\'ia y Ciencias, Universidad Diego Portales, Av. Ej\'ercito Libertador 441, Santiago, Chile\\
$^{9}$ Departamento de F\'{\i}sica de la Tierra y Astrof\'{\i}sica, Facultad de CC F\'{\i}sicas, Universidad Complutense de Madrid, E\mbox{-}28040 Madrid, Spain\\
$^{10}$ Instituto de F\'{\i}sica de Part\'{\i}culas y del Cosmos (IPARCOS), Facultad de CC F\'{\i}sicas, Universidad Complutense de Madrid, E\mbox{-}28040 Madrid, Spain\\
$^{11}$ Cahill Center for Astrophysics, California Institute of Technology, 1216 East California Boulevard, Pasadena, CA 91125, USA\\
$^{12}$ Max-Planck-Institut f\"ur Extraterrestrische Physik, Gie{\ss}enbachstra{\ss}e 1, 85748 Garching, Germany\\
$^{13}$ Max Planck Institute for Extraterrestrial Physics (MPE), Giessenbachstr.\ 1, 85748 Garching, Germany\\
$^{14}$ Department of Physics \& Astronomy, University of Alaska Anchorage, Anchorage, AK 99508\mbox{-}4664, USA\\
$^{15}$ Department of Physics, University of Alaska, Fairbanks, AK 99775\mbox{-}5920, USA\\
$^{16}$ Institute of Astrophysics, Foundation for Research and Technology\mbox{--}Hellas (FORTH), Heraklion, GR\mbox{-}70013, Greece\\
$^{17}$ School of Sciences, European University Cyprus, Diogenes Street, Engomi, 1516 Nicosia, Cyprus\\
$^{18}$ Observatorio Astron\'omico Nacional (OAN\mbox{--}IGN), Observatorio de Madrid, Alfonso XII 3, 28014 Madrid, Spain\\
$^{19}$ Department of Physics \& Astronomy, University of Southampton, Highfield, Southampton SO17 1BJ, UK\\
$^{20}$ Instituto de Radioastronom\'{\i}a y Astrof\'{\i}sica (IRyA), Universidad Nacional Aut\'onoma de M\'exico, Antigua Carretera a P\'atzcuaro \#8701, Colonia ExHda.\ San Jos\'e de la Huerta, Morelia, Michoac\'an, M\'exico C.P.\ 58089\\
$^{21}$ Department of Physics and Astronomy, The University of Texas at San Antonio (UTSA), 1 UTSA Circle, San Antonio, TX 78249\mbox{-}0600, USA\\
$^{22}$ Telespazio UK for ESA, ESAC, Camino Bajo del Castillo s/n, 28692 Villanueva de la Ca\~nada, Spain\\
$^{23}$ Space Telescope Science Institute, 3700 San Martin Drive, Baltimore, MD 21218, USA\\
$^{24}$ Department of Physics \& Astronomy, University of South Carolina, Columbia, SC 29208, USA\\
$^{25}$ Kavli Institute for Particle Astrophysics \& Cosmology (KIPAC), Stanford University, Stanford, CA 94305, USA\\
$^{26}$ National Astronomical Observatory of Japan, National Institutes of Natural Sciences (NINS), 2\mbox{-}21\mbox{-}1 Osawa, Mitaka, Tokyo 181\mbox{-}8588, Japan\\
$^{27}$ Instituto de F\'{\i}sica Fundamental (IFF), CSIC, Calle Serrano 123, 28006 Madrid, Spain\\
$^{28}$ Centro de Astrobiolog\'{\i}a (CAB), CSIC\mbox{--}INTA, Ctra.\ de Ajalvir km 4, Torrej\'on de Ardoz, E\mbox{-}28850, Madrid, Spain\\
$^{29}$ Departamento de F\'{\i}sica, CCNE, Universidade Federal de Santa Maria, Av.\ Roraima 1000, 97105\mbox{-}900, Santa Maria, RS, Brazil\\
$^{30}$ Laborat\'orio Nacional de Astrof\'{\i}sica (LNA/MCTI), 37530\mbox{-}000, Itajub\'a, MG, Brazil\\
$^{31}$ Observat\'orio Nacional, Rua General Jos\'e Cristino 77, 20921\mbox{-}400 S\~ao Crist\'ov\~ao, Rio de Janeiro, RJ, Brazil\\
$^{32}$ Astronomical Observatory, Volgina 7, 11060 Belgrade, Serbia\\
$^{33}$ Sterrenkundig Observatorium, Universiteit Gent, Krijgslaan 281\mbox{-}S9, B\mbox{-}9000 Gent, Belgium\\
$^{34}$ Centre for Extragalactic Astronomy, Department of Physics, Durham University, South Road, Durham, DH1 3LE, UK\\



\appendix

\section{NGC 2992: Warm vs Cold}
\label{ap:ngc2992}
\begin{figure*}
  \centering
  \includegraphics[width=0.7\textwidth]{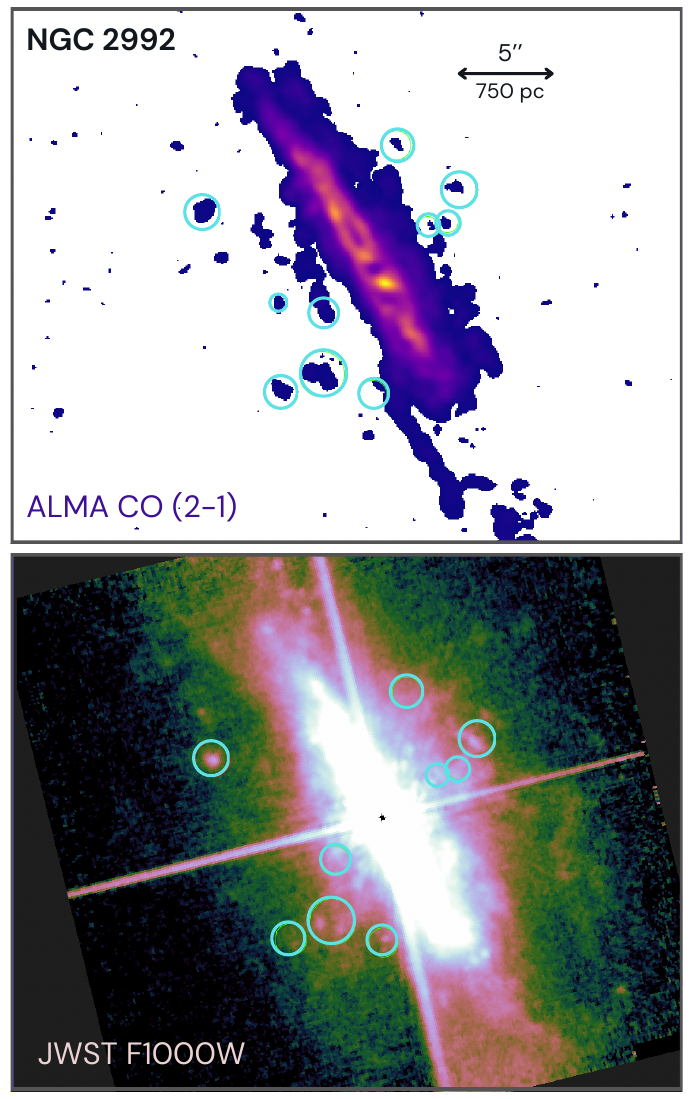}
  \caption{Comparison between the ALMA CO(2–1) flux map (courtesy of M. Zanchettin) and the JWST/MIRI F1000W image (bottom) of NGC 2992, in logarithmic stretch. North is up and east to the left. Cyan circles mark clumps identified in CO(2–1) by \citet{2023A&A...679A..88Z} (see their Fig. 6), several of which coincide with MIR clumps detected in the JWST image.}
  \label{NGC2992_ALMA_JWST}
\end{figure*}

\citet{2023A&A...679A..88Z} report on the detection of cold molecular clumps in NGC 2992 that are entrained with the outflow, with projected distance up to 1.7 kpc and velocities close to $\sim$ 200 km/s. In Fig.~\ref{NGC2992_ALMA_JWST}, we show that several of these clumps can also be detected in the MIR, albeit very faint.  The JWST and ALMA images are astrometrically aligned, so the positions of these 
clumps in the JWST image are reliable, although they are of very low surface brightness 
and only visible under strong stretching of the image. We do not convolve the JWST image to 
the ALMA beam, as this would wash out the already very faint dusty 
clumps and prevent a visual inspection.  The clumps are distributed along the NLR and also extend to kpcs in scale away from the nucleus. In agreement with  \citet{2023A&A...679A..88Z}, we find no spatial correspondence between these clumps and the radio bubbles, which are too compact and fall below the location of these clumps.  \citet{2023A&A...679A..88Z} propose that these clumps are likely linked to a previous AGN episode, which could explain their faint detection in JWST. 

\section{SEDs}
\begin{figure*}
  \centering
  \includegraphics[width=0.99\textwidth]{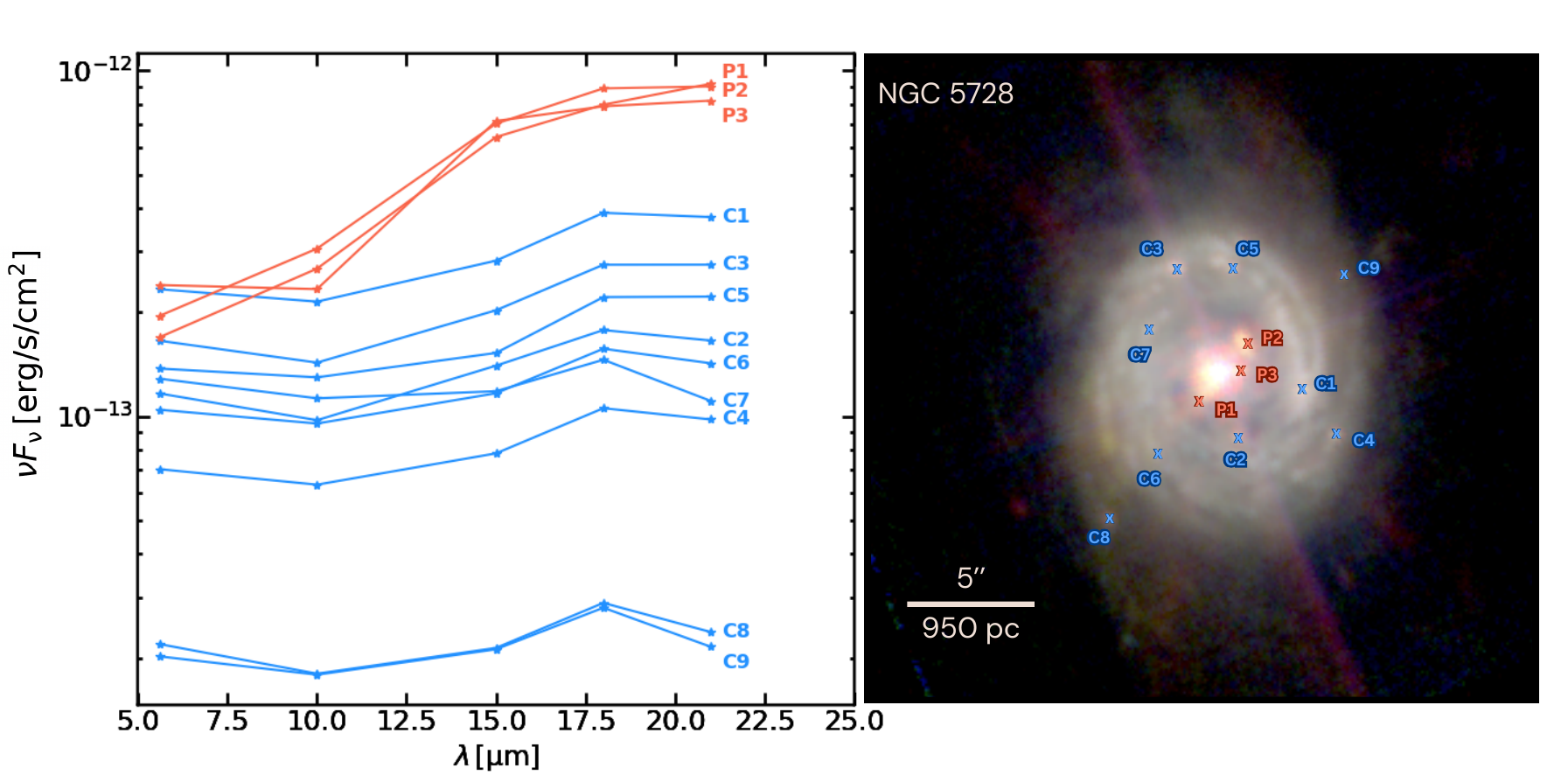}
  \caption{\textit{Left:} Spectral energy distributions  extracted from individual apertures on the PSF-subtracted and decontaminated images in NGC~5728. Red curves correspond to regions covering emission along the NLR, while blue curves  correspond to SF clumps in the disk. \textit{Right:} Illustrative JWST/MIRI three-colour image (R:F560W, G:F1000W, B:F1500W), produced from the original (non-PSF-subtracted) images, with ROIs overlaid to indicate the regions used for SED extraction.}
  \label{SEDs_NGC5728}
\end{figure*}

\begin{figure*}
  \centering
  \includegraphics[width=0.99\textwidth]{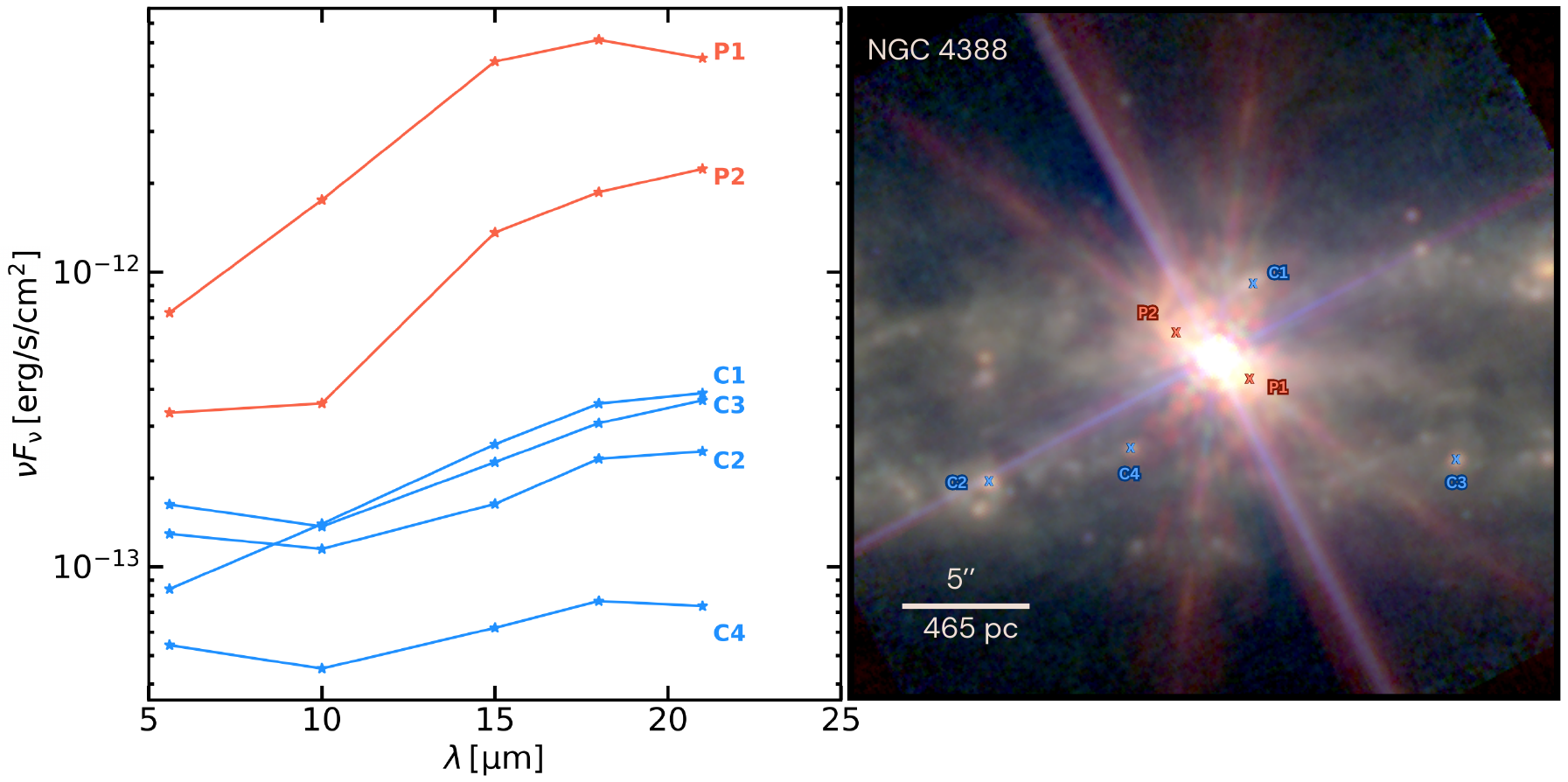}
  \caption{Same as Fig.~\ref{SEDs_NGC5728} but for NGC 4388.}
  \label{SEDs_NGC4388}
\end{figure*}

\begin{figure*}
  \centering
  \includegraphics[width=0.99\textwidth]{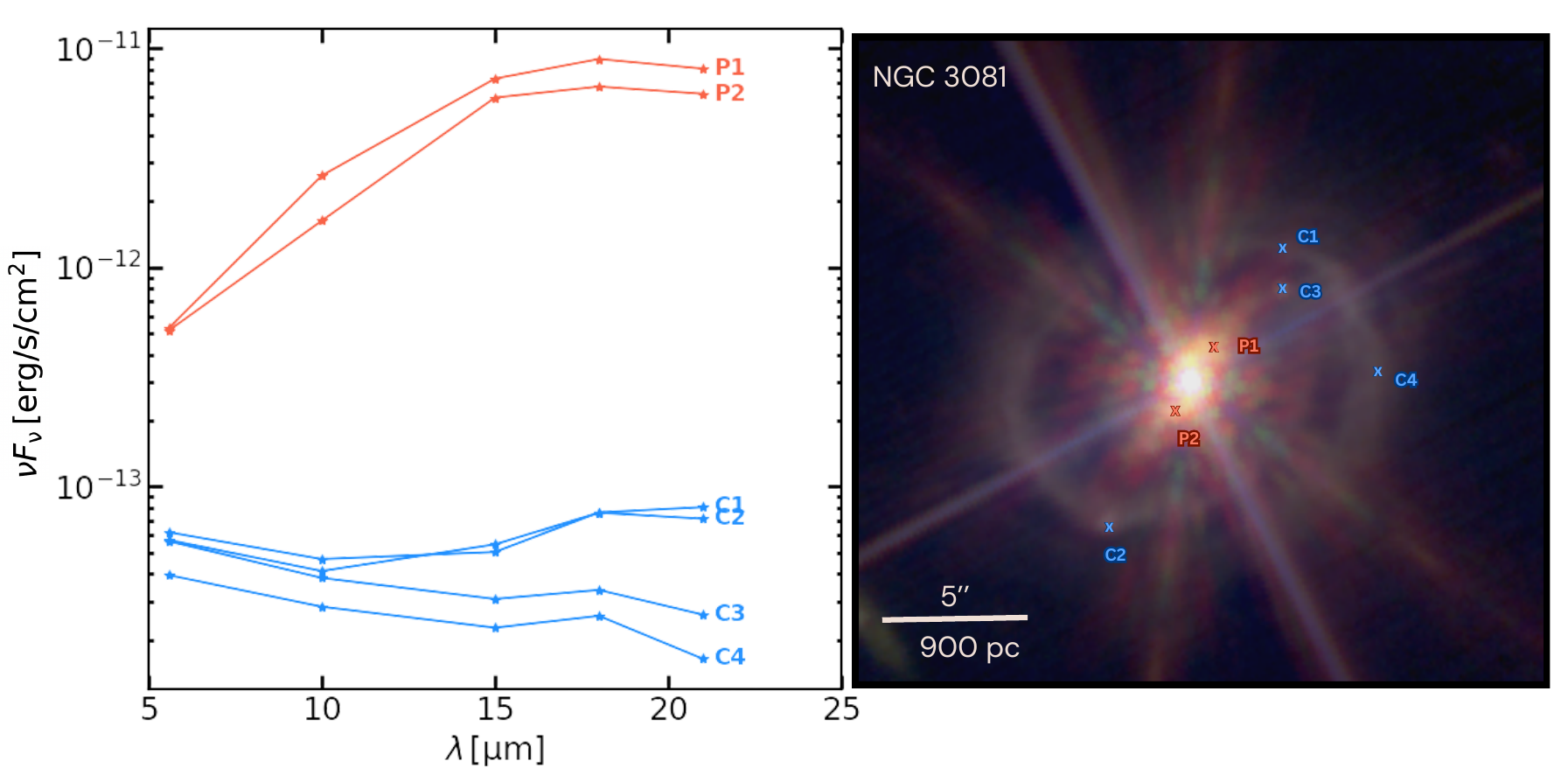}
  \caption{Same as Fig.~\ref{SEDs_NGC5728} but for NGC 3081.}
  \label{SEDs_NGC3081}
\end{figure*}

\begin{figure*}
  \centering
  \includegraphics[width=0.99\textwidth]{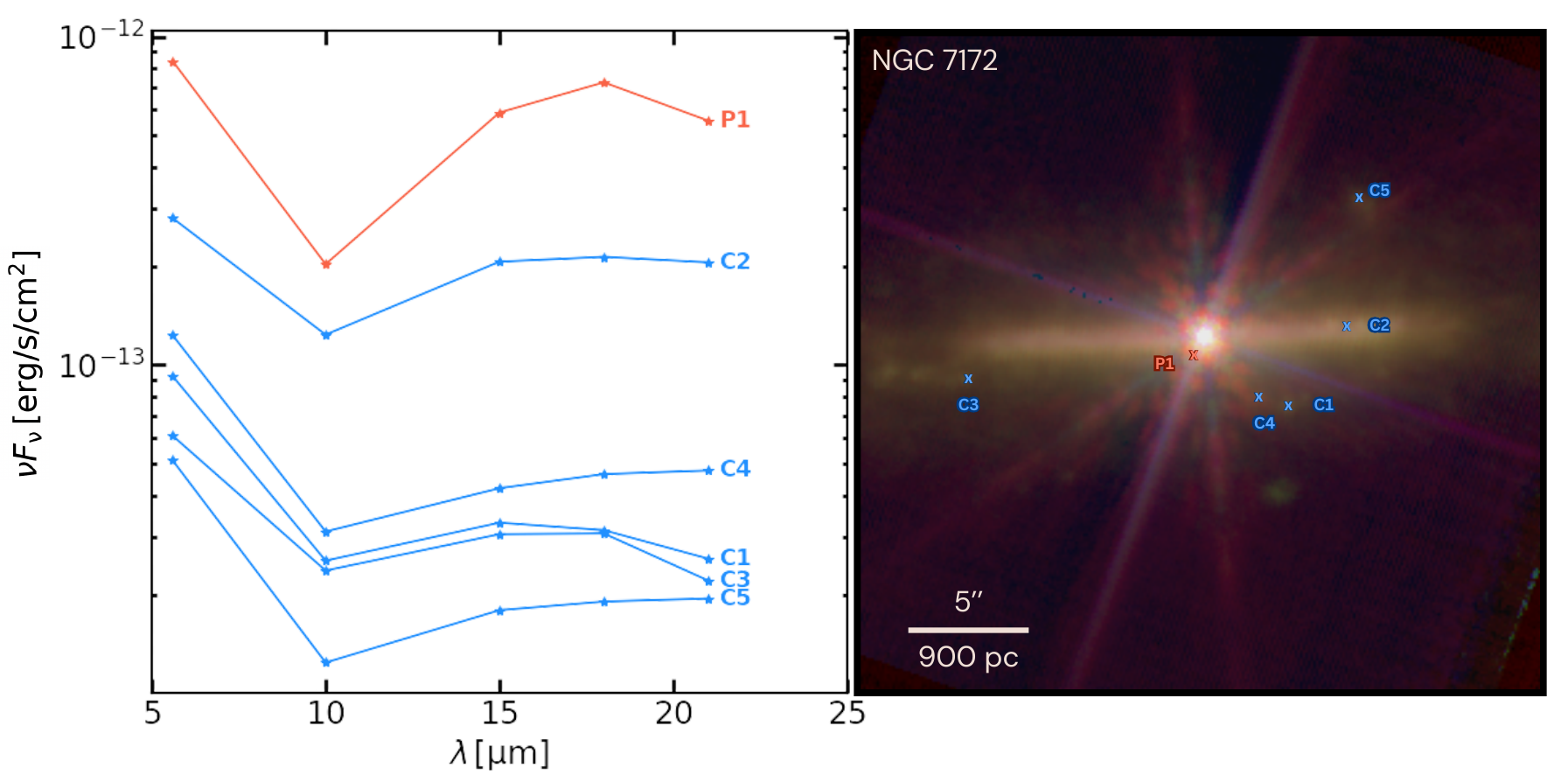}
  \caption{Same as Fig.~\ref{SEDs_NGC5728} but for NGC 7172.}
  \label{SEDs_NGC7172}
\end{figure*}

\begin{figure*}
  \centering
  \includegraphics[width=0.99\textwidth]{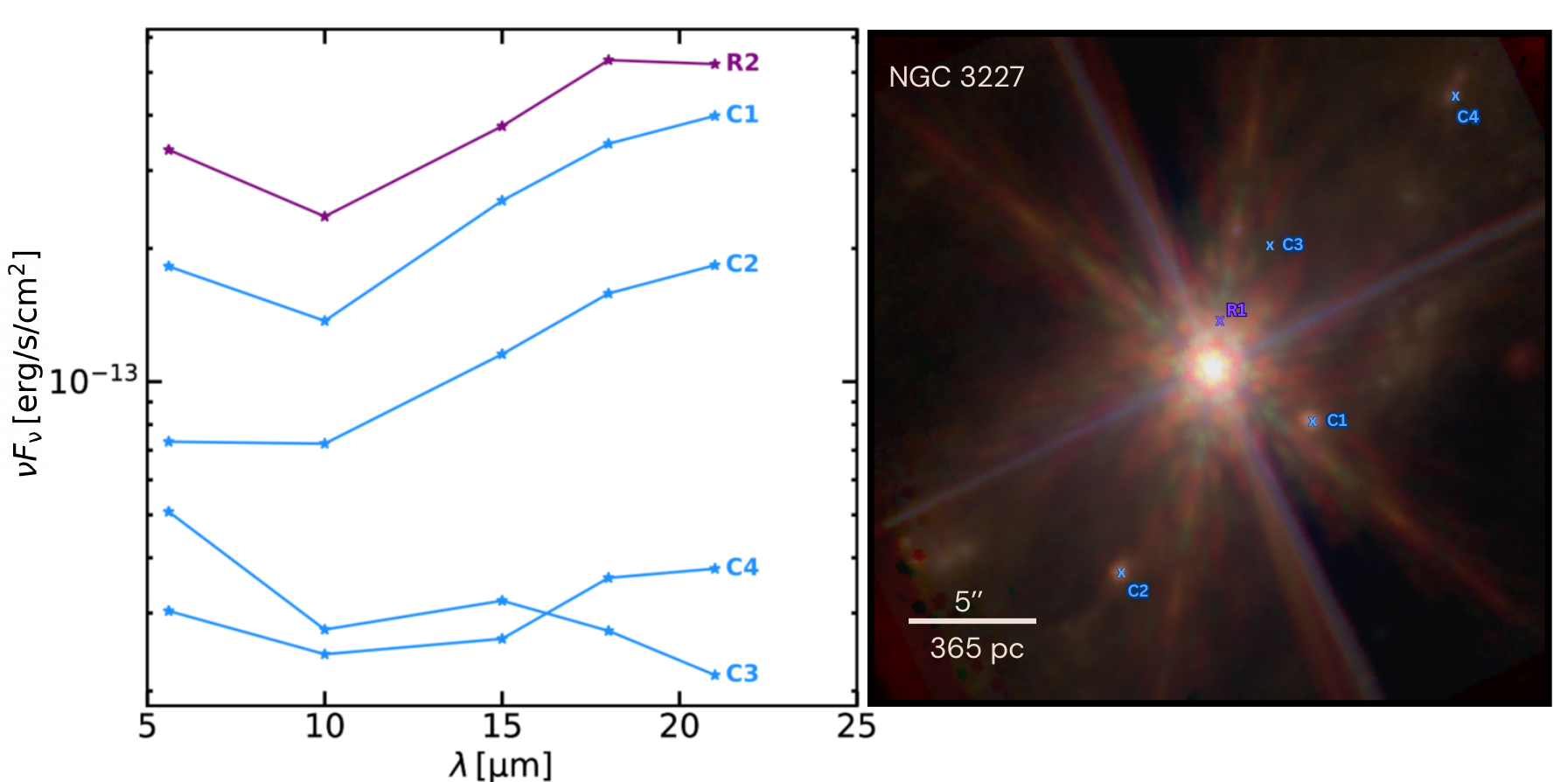}
  \caption{Same as Fig.~\ref{SEDs_NGC5728} but for NGC 3227. }
  \label{SEDs_NGC3227}
\end{figure*}

\begin{figure*}
  \centering
  \includegraphics[width=0.99\textwidth]{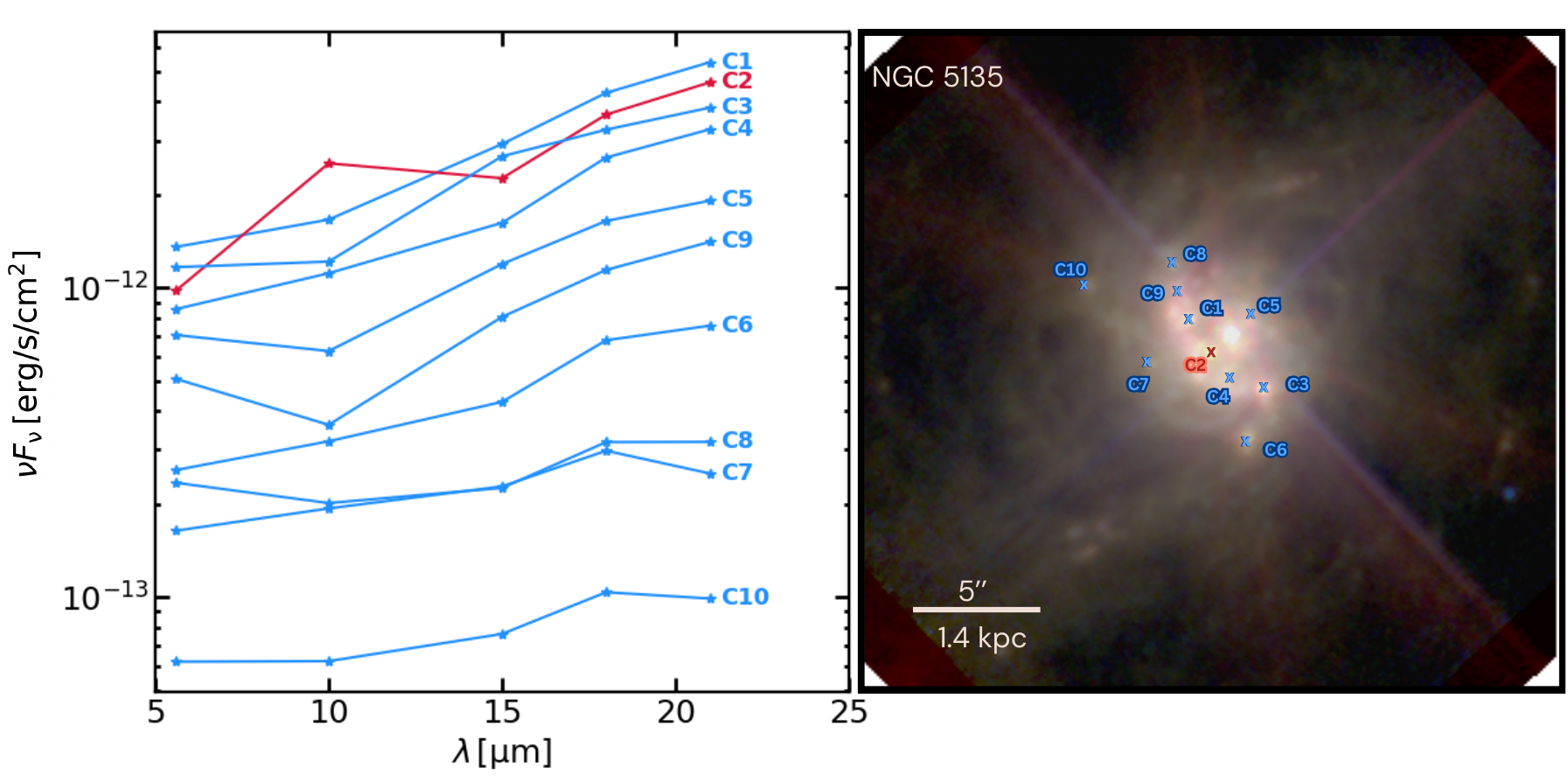}
  \caption{Same as Fig.~\ref{SEDs_NGC5728} but for NGC 5135.}
  \label{SEDs_NGC5135}
\end{figure*}

\begin{figure*}
  \centering
  \includegraphics[width=0.99\textwidth]{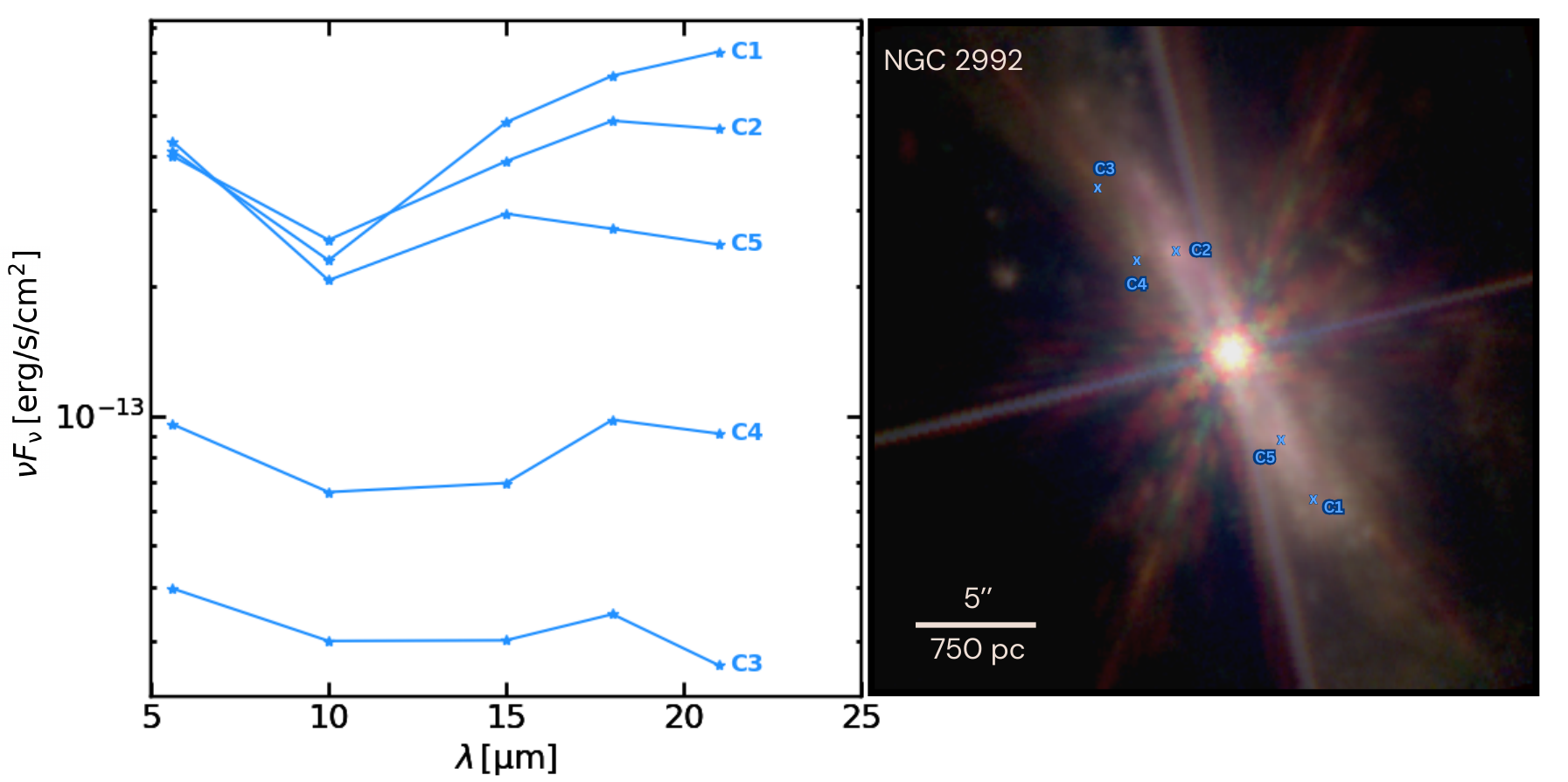}
  \caption{Same as Fig.~\ref{SEDs_NGC5728} but for NGC 2992.}
  \label{SEDs_NGC2992}
\end{figure*}

\begin{figure*}
  \centering
  \includegraphics[width=0.99\textwidth]{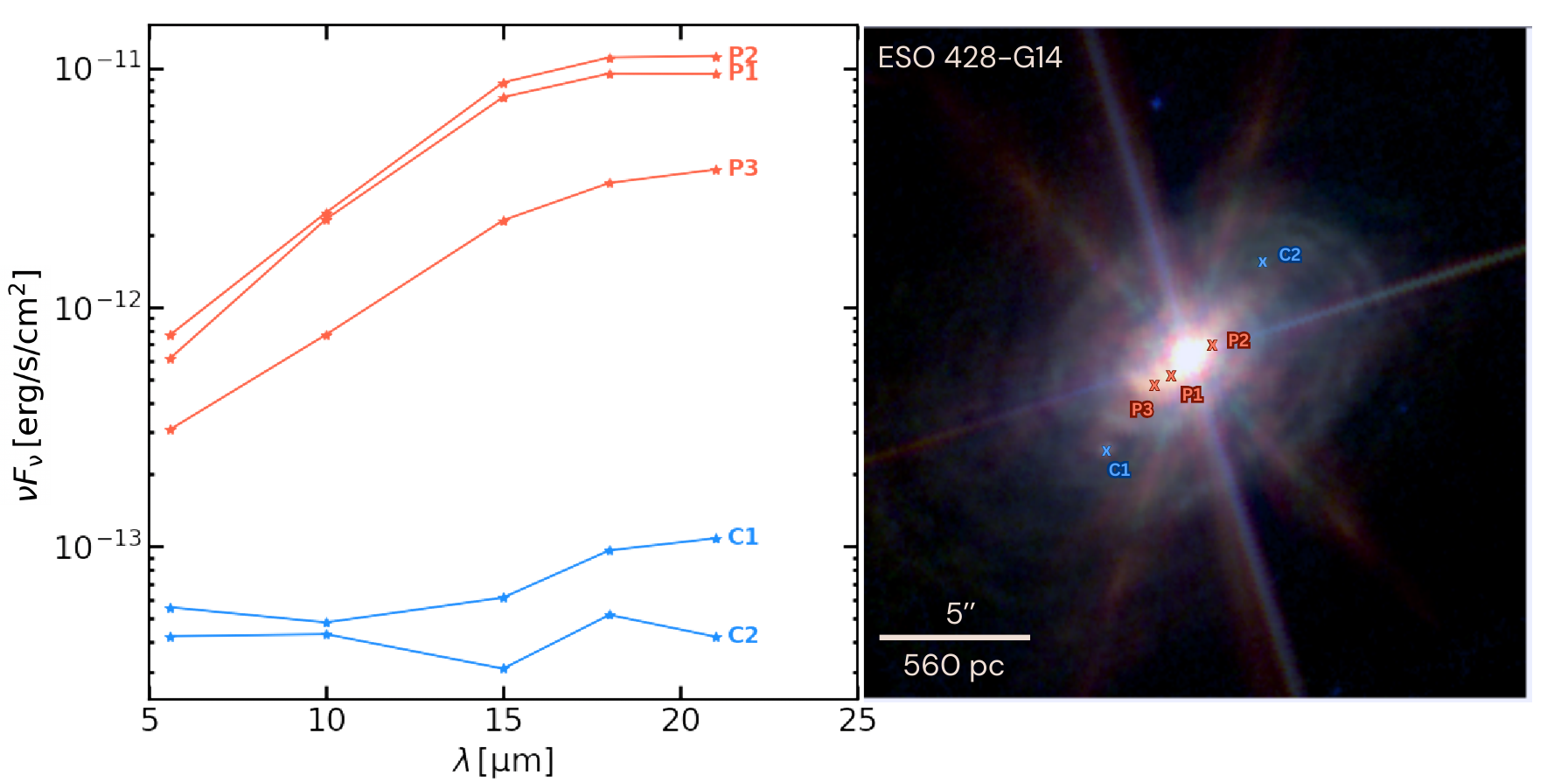}
  \caption{Same as Fig.~\ref{SEDs_NGC5728} but for ESO428-G14, also presented in Fig. 1 and 3 in \citet{Haidar+24}.}
  \label{SEDs_ESO428}
\end{figure*}

We extract SEDs across various ROIs and for all galaxies as presented in Figs.~\ref{SEDs_NGC5728}-~\ref{SEDs_ESO428}. For ESO 428$-$G14, the SEDs are presented and extensively studied in \citet{Haidar+24} (Fig. 3 therein). A case by case study of these SEDs is beyond the scope of this paper. Here 
we show them only for illustrative purposes to help interpret the medians presented in Fig.~\ref{Median_SEDs_allgals_polar_vs_SF}. Upcoming focused target studies from the GATOS programme will provide detailed case by case analyses of these SEDs.

\section{DUST-RADIO LINK FOR THE REST OF THE SAMPLE}
\label{ap:dust-radio-extra}
Just as done in Fig.~\ref{fig:all_radio_dust}, we compare the morphology of the dust with the radio and coronal line emissions for galaxies that show weak coupling between the dust and the ionised NLR (NGC 5135, NGC 7172, NGC 3227 and NGC 2992). We find that, on larger scales, the radio emission generally traces well the dust, except for NGC 2992, for which 4.8 GHz emission is dominated by the radio bubbles. Other interesting features include the SN shock in NGC 5135 \citep[][their Fig. 1]{2012ApJ...749..116C}, which appears to be co-spatial with dust in the galaxy disk. \citet{2012ApJ...749..116C} estimates that the total energy produced in SN shock is $\sim 2.7\times 10^{42}\, \rm erg/s$. This highlights the strength of shocks in this region and their potential role in shaping the dust in the ISM. NGC 3227 shows some extended radio emission that may be associate to the nuclear ring, which also shines bright in the MIR. \citet{2019MNRAS.485.2054S} found that this ring is shocked with high [\ion{Fe}{II}]/Pa$\beta$ ratios ($\ge 4$) throughout the ring (see their Fig. 4). This dust-shock relation, for both NCG 5135 and NGC 3227 further supports that dust can survive in these regions. 

\begin{figure*}
  \centering
  {\includegraphics[width=0.72\textwidth]{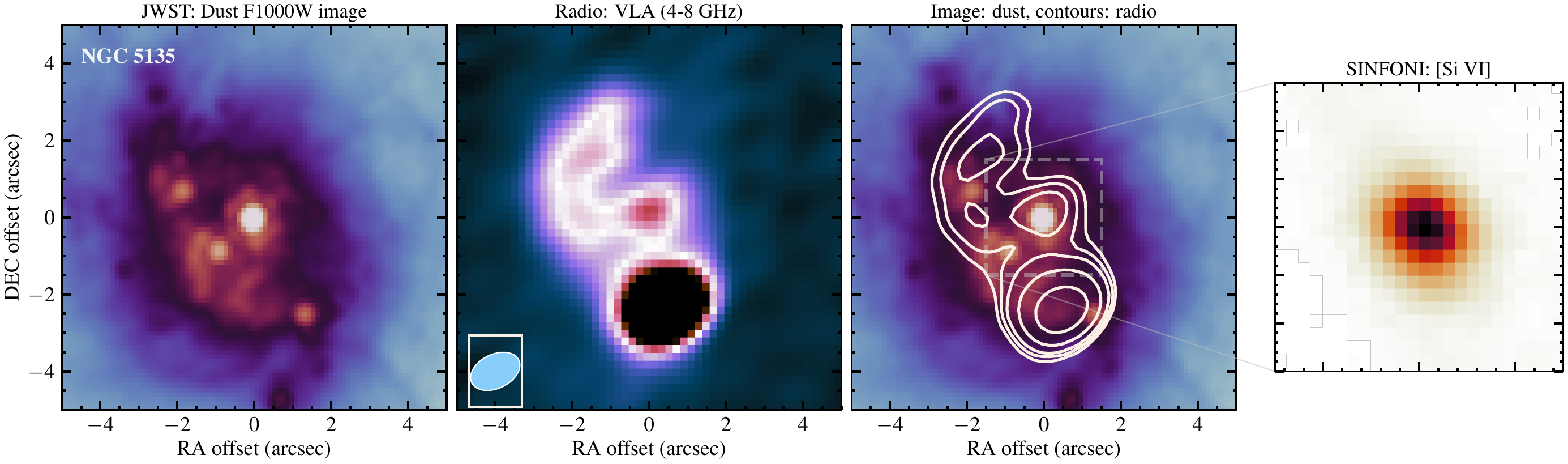}}
  {\includegraphics[width=0.73\textwidth]{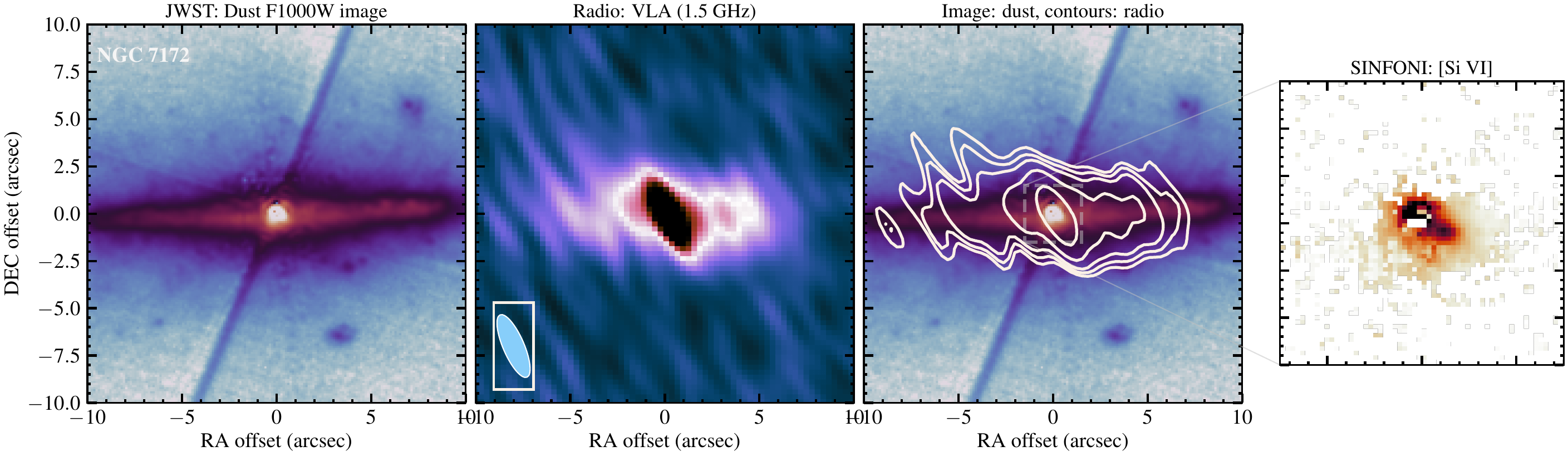}} \\
  {\includegraphics[width=0.72\textwidth]{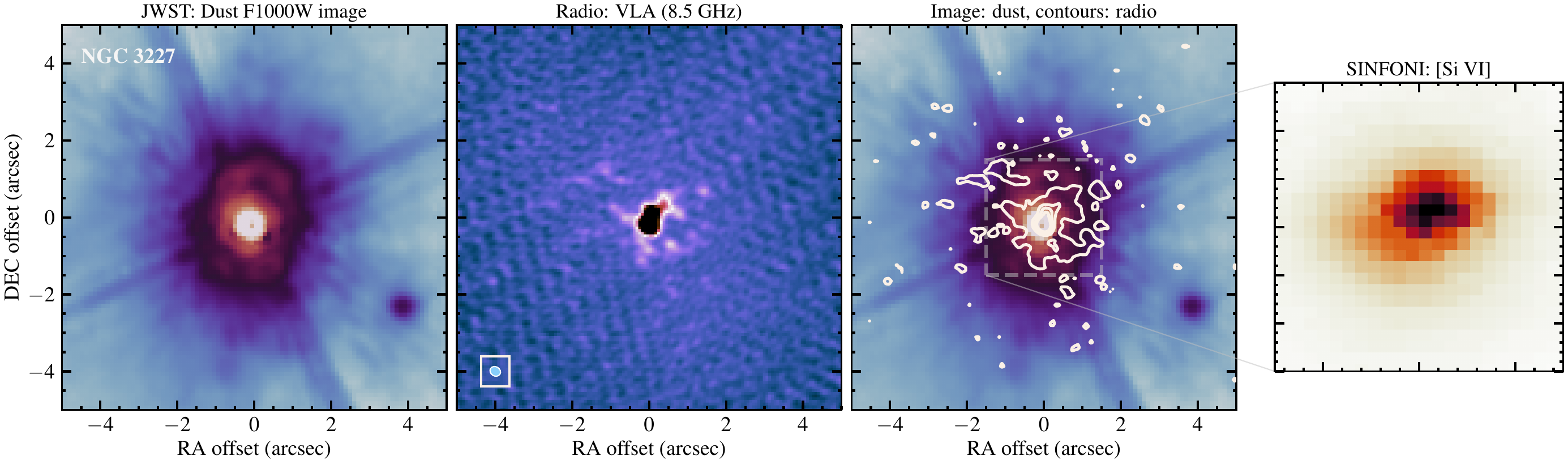}}
  {\includegraphics[width=0.72\textwidth]{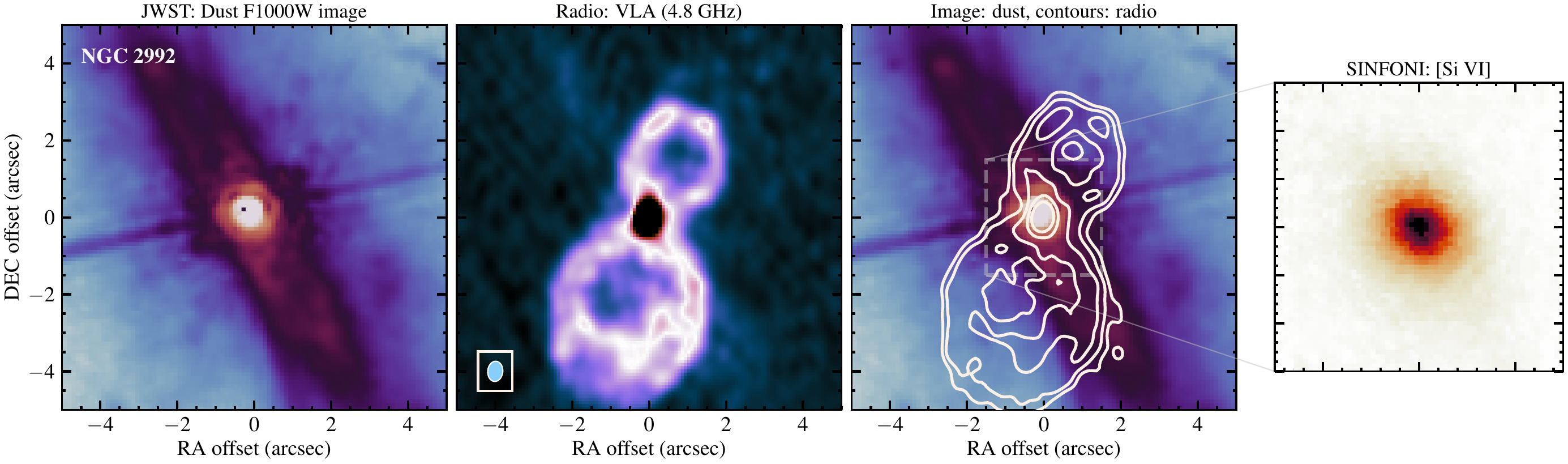}} \\

  \caption{Same as Fig.~\ref{fig:all_radio_dust}, but for the remaining four galaxies in the sample.}
  \label{fig:dust_radio_no_coupling}
\end{figure*}

\bsp	
\label{lastpage}
\end{document}